# Performance Evaluation of Equal-Weight Portfolio and Optimum Risk Portfolio on Indian Stocks

Abhiraj Sen[1] and Jaydip Sen[2]
[1]Jadavpur University, Kolkata, India, [2]Praxis Business School, Kolkata, India
email: {[1]abhirajsen@ieee.org, [2]jaydip.sen@acm.org}

**Abstract:** Designing an optimum portfolio for allocating suitable weights to its constituent assets so that the return and risk associated with the portfolio are optimized is a computationally hard problem. The seminal work of Markowitz that attempted to solve the problem by estimating the future returns of the stocks is found to perform sub-optimally on real-world stock market data. This is because the estimation task becomes extremely challenging due to the stochastic and volatile nature of stock prices. This work illustrates three approaches to portfolio design minimizing the risk, optimizing the risk, and assigning equal weights to the stocks of a portfolio. Thirteen critical sectors listed on the National Stock Exchange (NSE) of India are first chosen. Three portfolios are designed following the above approaches choosing the top ten stocks from each sector based on their free-float market capitalization. The portfolios are designed using the historical prices of the stocks from Jan 1, 2017, to Dec 31, 2022. The portfolios are evaluated on the stock price data from Jan 1, 2022, to Dec 31, 2022. The performances of the portfolios are compared, and the portfolio yielding the higher return for each sector is identified.

## 1. Introduction

Portfolio optimization is the process of identification of a set of capital assets and their respective weights allocation such that the risk-return pairs are optimized. The optimization problem is further confounded due to the requirement of the estimation of the future returns of stocks that exhibit volatility in their prices. Quite a few experts on the *efficient market hypothesis* (Fama, 1998; Fama, 1970; Fama & French, 1998; Fama & French, 1993) believe that accurate forecasting of future stock prices is not possible. Numerous propositions in the literature, however, illustrate the effective usage of complex algorithms and the design of predictive models to precisely predict future stock prices. Several propositions have been made on various portfolio optimization approaches following Markowitz's seminal work on minimum-variance portfolios (Markowitz, 1952). Several statistical and econometric, and learning-based approaches for future stock price prediction based on methods such as *multivariate regression*, *autoregressive integrated moving average* (ARIMA), *vector autoregression* (VAR), time series forecasting, machine learning, and deep learning, have also been proposed in the literature.

This paper presents a methodical approach to building robust stock portfolios in thirteen key sectors of the Indian economy. For each of these thirteen sectors, the top ten stocks based on their free-float market capitalization in the national stock exchange (NSE) are identified (NSE Website). The ticker names of these 130 stocks are used to automatically scrape their historical prices from the Yahoo Finance website (Yahoo Finance Website). Using these historical prices over five years, three different portfolios are designed following the equal weight allocation, minimum risk portfolio allocation, and mean-variance portfolio allocation. In the rest of the chapter, the mean-variance portfolio is referred to as the optimum risk portfolio. The portfolio performances are evaluated based on their annual returns over the year 2022. Moreover, other characteristics of the portfolios such as risk, weights assigned to constituent stocks, and the correlation among the stocks, are also analyzed. The study elucidates a comparative understanding of portfolio performances, along with actual return and volatility for the thirteen sectors. As an immediate use of this work, stock market investors can exploit the results of this work to gain comprehensive information on the current return on investment (ROI), and the risk involved in the sectors and the stocks analyzed in this work.

The work has three major contributions. First, this work presents two approaches toward stock portfolio design, the equal-weight portfolio, and the mean-variance portfolio. These two portfolio design approaches are used to build sectoral portfolios of stocks from thirteen important sectors listed in the NSE, India. The performance results of these portfolios will be a good indicator of the current profitability of these sectors, and hence, the results will be a good guide for investors of the Indian stock market. Second, the results of this work will enable one to understand the potential return and the risk associated with the stocks of the thirteen sectors analyzed in this study. Third, the study also reveals which portfolio design approach yields a higher return and a lower risk for a given sector.

The paper is organized as follows. Section 2 presents a summary of some of the important related works. Section 3 describes the data and methodology pursued in the design of portfolios. Section 4 provides extensive results on portfolio performances, along with a detailed analysis of the same. Finally, the paper is concluded in Section 5, in which some future directions of work are highlighted.

## 2. Related Work

Portfolio design and optimization is a challenging problem for which numerous solutions and approaches have been proposed by researchers. Portfolio design and optimization is a challenging problem that has attracted considerable attention from researchers. Numerous approaches have been proposed to solve this complex problem involving robust stock price prediction and the formation of the optimized combination of stocks to maximize the return on investment. Machine learning models have been extensively used by researchers in predicting future stock prices (Carta et al., 2021; Chatterjee et al., 2021; Mehtab & Sen, 2021; Mehtab & Sen, 2020a; Mehtab & Sen, 2019; Mehtab et al., 2021; Sarmento & Horta, 2020; Sen, J., 2018a; Sen & Datta Chaudhuri, 2017a). The prediction accuracies of the models are found to have been improved by the use of deep learning architectures and algorithms (Chatterjee et al., 2021; Chen et al, 2018; Chong et al., 2017; Mehtab & Sen, 2022; Mehtab & Sen, 2021; Mehtab & Sen, 2020a; Mehtab & Sen, 2020b; Mehtab & Sen, 2019; Mehtab et al., 2021; Mehtab, et al., 2020; Sen, 2018a; Sen & Mehtab, 2021a; Sen & Mehtab, 2021b; Sen et al., 2021a; Sen et al., 2021b; Sen et al., 2021i; Sen et al., 2020, Thormann et al., 2021; Tran et al., 2019). Several approaches to text mining have been effectively applied on social media and the web to improve prediction accuracies even further (Li & Pan, 2022; Mehtab & Sen, 2019; Thormann et al., 2021; Zhang et al., 2021). Among the other alternative approaches for stock price prediction, time series decomposition-based statistical and econometric approaches are also quite popular (Chatterjee et al., 2021; Cheng et al., 2018; Sen, 2022a; Sen, 2018b; Sen, 2017a; Sen, 2017b; Sen & Datta Chaudhuri, 2018; Sen & Datta Chaudhuri, 2017b; Sen & Datta Chaudhuri, 2017c; Sen & Datta Chaudhuri, 2016a; Sen & Datta Chaudhuri, 2016b; Sen & Datta Chaudhuri, 2016c; Sen & Datta Chaudhuri, 2016d; Sen & Datta Chaudhuri, 2015). For estimating the future volatility and risk of stock portfolios the use of several variants of GARCH has been proposed in some works (Sen et al., 2021d). Over the last few years, reinforcement learning has been extensively used in robust and accurate prediction of stock prices and portfolio design (Brim, 2020; Fengqian & Chao, 2020; Kim et al., 2022; Kim & Kim, 2019; Lei et al., 2020; Li et al., 2019; Lu et al., 2021; Park & Lee, 2021).

The classical mean-variance optimization approach is the most well-known method for portfolio optimization (Sen & Mehtab, 2022; Sen & Dutta, 2022b; Sen et al., 2021e; Sen et al., 2021g; Sen et al., 2021h). Several alternatives to the mean-variance approach to portfolio optimization have also been proposed by some researchers. Notable among these methods are multiobjective optimization (Wang et al, 2022; Zheng & Zheng, 2022), Eigen portfolios using principal component analysis (Sen & Dutta, 2022a; Sen & Mehtab, 2022), risk parity-based methods (Sen & Dutta, 2022a; Sen & Dutta, 2021; Sen et al., 2021c; Sen et al., 2021f), and swarm intelligence-based approaches (Corazza et al., 2021; Thakkar & Chaudhuri, 2021). The use of genetic algorithms (Kaucic et al., 2019), fuzzy sets (Karimi et al., 2022), prospect theory (Li et al., 2021), and quantum evolutionary algorithms (Chou et al., 2021), cointegration-based approaches (Sen, 2022b; Sen, 2023c) are also proposed in the literature.

In the present work, the design of portfolios of stocks in thirteen key sectors of the Indian stock market is done following three portfolio-building methodologies, viz., equal weight portfolio, minimum risk portfolio, and optimum risk portfolio. The portfolios for each of these sectors are designed using the historical prices of the top ten stocks in NSE over five years (i.e., from Jan 1, 2017, to Dec 31, 2021). For each sector, the performances of the equal-weight portfolio and the mean-variance portfolio are evaluated based on their annual returns over the test period from Jan 1, 2022, to Dec 31, 2022, and the portfolio yielding the higher return is identified.

## 3. Data and Methodology

This section presents a detailed discussion of the steps involved in the data acquisition, data preprocessing, and the design of the portfolios for the thirteen sectors of stocks listed in the NSE. The process consists of seven steps which are as follows.

### 3.1. The selection of the sectors

As discussed in Section 1, thirteen sectors are selected from the NSE, India. The sectors are *auto, banking, consumer durables, financial services, fast-moving consumer goods* (FMCG)*, information technology* (IT)*, media, metal, oil & gas, pharma, public sector banks, private banks,* and *realty*. Based on the free-float market capitalization in the NSE, the top ten stocks for each of the abovementioned sectors are identified. The sector-specific portfolios are built using those stocks.

## 3.2. Data acquisition

For each sector, the prices of the stocks are acquired from the Yahoo Finance site using the *YahooFinancials* function. The parameters passed in the function are the *ticker name*, and the *date range* from Jan 1, 2017, to Dec 31, 2022, with "daily" frequency. The extracted stock data have the following attributes:

 i. *open*
 ii. *high*
 iii. *close*
 iv. *volume*
 v. *adjusted close*

This work is based on univariate analysis and hence the variable 'Close' is chosen as the variable of interest, while the remaining variables are ignored. The close values of the ten stocks, from Jan 1, 2017, to Dec 31, 2021, are used for building the portfolios. The portfolios are tested for their returns from Jan 1, 2022, to Dec 30, 2022.

## 3.3. Deriving the stock returns and volatility

The *daily return*, which is the percentage change in consecutive daily *close* prices, for each stock is computed on the training data using the *pct_change* function of Python. The annual return for each stock is computed as the weighted sum of the daily returns for the stock. Using the daily return values, the daily and annual volatility of the stocks are then calculated, using (1) and (2). ions.

$$Daily\ volatility\ (D_v) = standard\ deviation\ of\ daily\ returns \qquad (1)$$

$$Annual\ volatility = D_v \times \sqrt{250}, assuming\ 250\ working\ days/year\ for\ stock\ market \qquad (2)$$

Since annual volatility indicates price variability, it provides investors with a measure of the risk associated with a stock. The *std* function of Python function is used to calculate the volatility of stocks.

## 3.4. The covariance and correlation matrices computation

In the next step, to get an understanding of the association strength between the close prices of a pair of stocks in a sector, their covariance and correlation matrices are calculated using the training dataset. A strong association of a pair is indicated by a high covariance value between them and vice-versa. The matrices are computed using the *cov* and *corr* Python functions. A good portfolio aims to optimize the return while minimizing its risk. To minimize the risk, the identification of stocks with low correlation among them is required. This helps in achieving a higher diversity in the portfolio, thereby reducing its risk.

## 3.5. The design of equal-weight portfolios

Now we delve deep into historical price analysis of the chosen stocks in each sector. Using equal weights of the stocks, sector-wise portfolios are created. As we are working with ten stocks per sector, the weight assigned to every stock is 0.1. The yearly returns, along with associated risks, are calculated based on the training data assuming equal weight allocation to the stocks. The expected return of a portfolio is computed using (3) in which, *E(R)* denotes the expected return of an *n*-stock portfolio where each stock is denoted as $S_1, S_2, \ldots S_n$. The associated weights of the stocks are represented by $w_i$'s.

$$E(R) = w_1 E(R_{S_1}) + w_2 E(R_{S_2}) + \cdots + w_n E(R_{S_n}) \tag{3}$$

For calculating the equal weights portfolios of each sector, the *resample* function of Python is used with the parameter set to 'Y' so that the yearly mean values are computed.

### 3.6 The design of the minimum risk portfolios

Similar to the equal weight portfolios, the minimum risk portfolios are designed for all sectors. A portfolio with the minimum variance is identified as the minimum risk portfolio. The variance of a portfolio is calculated using variances of every stock and covariances between each stock-pair in it, as shown in (4).

$$Var(P) = \sum_{i=1}^{n} w_i \sigma_i^2 + 2 \sum_{i,j} w_i w_j Cov(i,j) \tag{4}$$

Since 10 stocks of each sector are used in this work, the variance computation involves 55 terms per portfolio, where the weighted variances contribute 10 terms and the weighted covariances contribute the remaining 45. To build the minimum risk portfolios, we find the weight combination that minimizes the portfolio variance. For each portfolio, first, the efficient frontier is plotted. The efficient frontier is the contour of the portfolios that indicates portfolios with the maximum returns for a given risk (Sen & Mehtab, 2022). The leftmost point on the efficient frontier indicates the portfolio with minimum risk. For plotting the effective frontier, 10000 portfolios are designed with random weights allocation to the 10 constituent stocks using a for loop in a Python program. Each of these 10000 portfolios is represented as a point in two-dimensional space, constituting the efficient frontier. That point on the efficient frontier, which is at the leftmost position, is used as a minimum-risk portfolio.

### 3.7 The design of the optimum risk portfolio

Investors are normally not interested in the low return values yielded by the minimum risk portfolio. Generally, they have the appetite to take some risks, provided the returns are high. A metric known as the Sharpe ratio is used to tradeoff portfolio return and risk. The Sharpe ratio of a portfolio is defined as the ratio of the return yielded by the portfolio to the return of a risk-free portfolio (Sen & Dutta, 2022b). This ratio is used to identify the portfolio with the optimum risk. In computing the Sharpe ratio, the volatility of the risk-free portfolio is assumed to be 1%. By maximizing the Sharpe ratio for the constituent stocks, the optimum risk portfolio trades off the risk with the return, thereby making the return significantly higher. The *idmax* function of Python has been used to find the portfolio with the highest Sharpe ratio (i.e., the optimum risk portfolio) among all points of an effective frontier.

## 4. Experimental Results

In this section, the details of the portfolio design, their performance results, and an extensive analysis of the results are presented. Python 3.7.4 and its libraries have been used to implement the portfolio models, while training and testing of the models have been done using Jupyter Notebook. A work station running on the Windows 11 operating system, an Intel i5-9300H CPU with a clock speed of 2.40 GHz, and a RAM of 8 GB have been used for training and testing of the portfolios. The results and analysis for each sector are presented separately as follows.

### 4.1 The auto sector portfolios

Following are the top ten stocks of this sector based on their free-float market capitalization in the NSE, as per the report released on Dec 30, 2022. The figures below represent the contributions in percentage, of the stocks (computed based on the market capitalization) to the overall index of the auto sector. The ticker name is also mentioned with a pair of parentheses just beside the name of the respective stock.

- i.   Mahindra & Mahindra (M&M) : 20.08
- ii.  Maruti Suzuki India (MARUTI) :18.74
- iii. Tata Motors (TATAMOTORS): 11.69
- iv.  Eicher Motors (EICHERMOT): 7.56
- v.   Bajaj Auto (BAJAJ-AUTO): 6.87
- vi.  Hero MotoCorp (HEROMOTOCO): 5.97
- vii. Tube Investments of India (TIINDIA): 4.86
- viii. TVS Motor Company (TVSMOTOR): 4.24
- ix.  Bharat Forge (BHARATFORG): 3.78
- x.   Ashok Leyland (ASHOKLEY): 3.46

**Table 1.** The return and risk of the auto sector stocks

| Stock | Annual Return (%) | Annual Risk (%) |
|---|---|---|
| M&M | 6.22 | 32.68 |
| MARUTI | -5.92 | 30.89 |
| TATAMOTORS | 27.22 | 47.92 |
| EICHERMOT | -2.91 | 33.53 |
| BAJAJ-AUTO | 0.29 | 26.38 |
| HEROMOTOCO | -8.21 | 29.98 |
| TIINDIA | 63.19 | 39.29 |
| TVSMOTOR | -2.77 | 33.86 |
| BHARATFORG | 1.52 | 38.17 |
| ASHOKLEY | 2.74 | 45.06 |

Table 1 represents the annual return and volatility for auto-sector stocks for the training period from Jan 1, 2017, to Dec 31, 2021. TIINDIA shows the highest annual return while MARUTI yields the lowest. TATAMOTORS has the highest annual risk while BAJAJ-AUTO has the lowest.

Table 2 presents the weights allocation to different stocks, using three portfolio design approaches:

1. Equal weight portfolio (EWP)
2. Minimum risk portfolio (MRP)
3. MVP portfolio (MVP) / Optimum risk portfolio (ORP)

In each of the above cases, the sum of weights is 1.

**Table 2.** The portfolio weights for the auto sector stocks

| Stock | EWP | MRP | MVP/ORP |
|---|---|---|---|
| M&M | 0.1 | 0.064107 | 0.016697 |
| MARUTI | 0.1 | 0.163289 | 0.023862 |
| TATAMOTORS | 0.1 | 0.028795 | 0.178039 |
| EICHERMOT | 0.1 | 0.122902 | 0.079096 |
| BAJAJ-AUTO | 0.1 | 0.242138 | 0.169046 |
| HEROMOTOCO | 0.1 | 0.072284 | 0.011963 |
| TIINDIA | 0.1 | 0.155767 | 0.372677 |
| TVSMOTOR | 0.1 | 0.149395 | 0.090742 |
| BHARATFORG | 0.1 | 0.000088 | 0.038361 |
| ASHOKLEY | 0.1 | 0.001234 | 0.019517 |

BAJAJ-AUTO receives the highest allocation as per the minimum risk portfolio while TIINDIA receives the highest allocation as per the optimum risk portfolio.

Figure 1 depicts the weight allocation to the auto sector stocks done by the optimum risk portfolio.

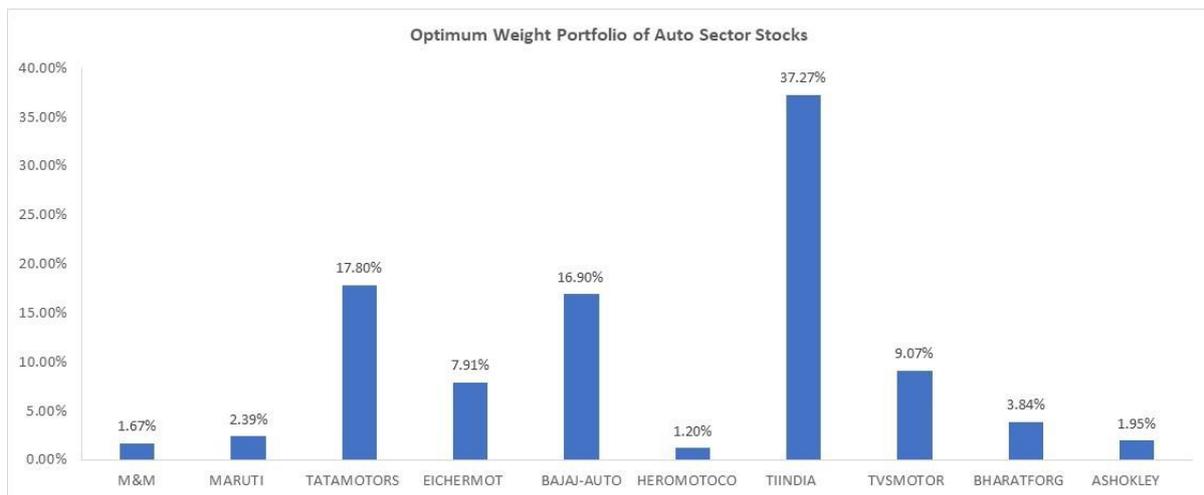

**Figure 1.** The ORP portfolio weights for the auto sector stocks

**Table 3.** The return and risk of the auto sector portfolios

| Metric | EWP | MRP | MVP/ORP |
|---|---|---|---|
| Portfolio annual return (%) | 8.14 | 8.77 | 27.94 |
| Portfolio annual risk (%) | 24.46 | 22.45 | 25.60 |

The risk and return values for the three portfolios, computed using stock prices over the training period, are depicted in Table 3 above. It is evident that the equal weights portfolio has yielded the lowest return and the minimum risk portfolio has exhibited the least risk. The optimum risk portfolio has produced the highest return and the highest risk. It should be noted that since the minimum risk portfolio yields a very low return, it is not adopted by investors. Hence, its performance is not evaluated over the test data for all sectors.

**Table 4.** The return of the equal-weight portfolio of the auto sector stocks

| Stock | Date: Jan 3, 2022 | | | Date: Dec 31, 2022 | | | RETURN |
|---|---|---|---|---|---|---|---|
| | Weights | Price / Stock | Amount Invested | No. of Stock | Price / Stock | Value of Stock | |
| M&M | 0.1 | 830 | 10000 | 12.05 | 1249 | 15054 | |
| MARUTI | 0.1 | 7524 | 10000 | 1.33 | 8395 | 11157 | |
| TATAMOTORS | 0.1 | 498 | 10000 | 20.1 | 388 | 7796 | |
| EICHERMOT | 0.1 | 2719 | 10000 | 3.68 | 3228 | 11872 | 23.52% |
| BAJAJ-AUTO | 0.1 | 3277 | 10000 | 3.05 | 3616 | 11034 | |
| HEROMOTOCO | 0.1 | 2477 | 10000 | 4.04 | 2739 | 11059 | |
| TIINDIA | 0.1 | 1889 | 10000 | 5.29 | 2776 | 14696 | |
| TVSMOTOR | 0.1 | 629 | 10000 | 15.89 | 1085 | 17249 | |
| BHARATFORG | 0.1 | 711 | 10000 | 14.07 | 880 | 12377 | |
| ASHOKLEY | 0.1 | 128 | 10000 | 78.31 | 143 | 11229 | |
| | | | 100000 | | | 123523 | |

The annual return for an investor, investing INR 100,000 on Jan 3, 2022, and following the equal weight portfolio approach is shown in Table 4. The investor receives a return of 23.52% at the end of the twelve months.

**Table 5.** The return of the optimum risk portfolio of the auto sector stocks

| Stock | Date: Jan 3, 2022 | | | Date: Dec 31, 2022 | | | RETURN |
|---|---|---|---|---|---|---|---|
| | Weights | Price | Amount Invested | No. of Stock | Price | Value of Stock | |
| M&M | 0.016697 | 830 | 1670 | 2.01 | 1249 | 2514 | |
| MARUTI | 0.023862 | 7524 | 2386 | 0.32 | 8395 | 2662 | |
| TATAMOTORS | 0.178039 | 498 | 17804 | 35.78 | 388 | 13881 | |
| EICHERMOT | 0.079096 | 2719 | 7910 | 2.91 | 3228 | 9390 | 25.78% |
| BAJAJ-AUTO | 0.169046 | 3277 | 16905 | 5.16 | 3616 | 18653 | |
| HEROMOTOCO | 0.011963 | 2477 | 1196 | 0.48 | 2739 | 1323 | |
| TIINDIA | 0.372677 | 1889 | 37268 | 19.73 | 2776 | 54770 | |
| TVSMOTOR | 0.090742 | 629 | 9074 | 14.42 | 1085 | 15652 | |
| BHARATFORG | 0.038361 | 711 | 3836 | 5.4 | 880 | 4748 | |
| ASHOKLEY | 0.019517 | 128 | 1952 | 15.28 | 143 | 2192 | |
| | | | 100000 | | | 125785 | |

The performance of the optimum risk portfolio for the auto-sector stocks is shown in Table 5. To compare the performance of this portfolio with that of the equal-weight portfolio, the initial amount of investment of INR 100,000 is kept constant. The return yielded by the optimum risk portfolio is found to be 25.78% as depicted in Table 5.

Finally, the efficient frontier for the auto sector portfolios is presented in Figure 2. In Figure 2, the minimum risk portfolio is denoted by the red star, and the optimum risk portfolio is denoted by the green star. The stock prices from Jan 1, 2017, to Dec 31, 2021, are considered in plotting the efficient frontier of the auto sector. It is to be noted that in Figure 2, the x-axis denotes the risk while the y-axis denotes the return.

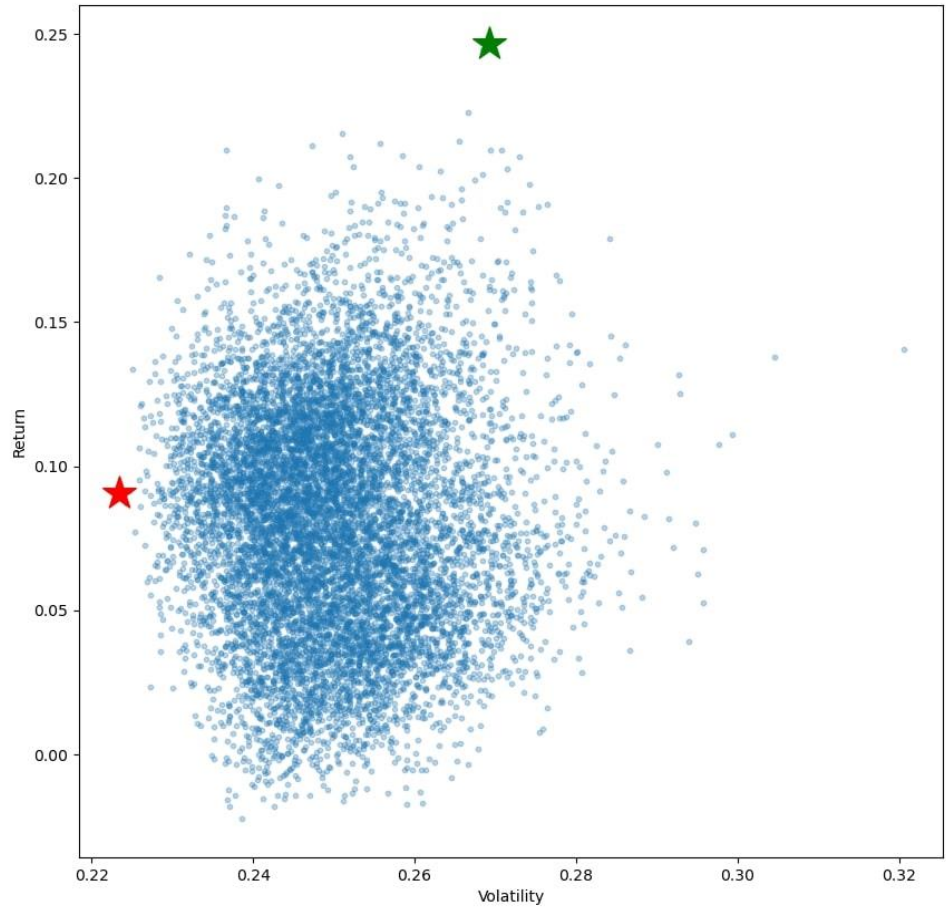

**Figure 2.** The efficient frontier of the auto sector portfolios

### 4.2 The banking sector portfolios

Following are the top ten stocks of this sector based on their free-float market capitalization in the NSE, as per the report released on Dec 30, 2022. The figures below represent the contributions in percentage, of the stocks (computed based on the market capitalization) to the overall index of the banking sector. The ticker name is also mentioned with a pair of parentheses just beside the name of the respective stock.

    i.     HDFC Bank (HDFCBANK): 28.66
    ii.    ICICI Bank (ICICIBANK):23.54
    iii.   Kotak Mahindra Bank (KOTAKBANK): 10.18
    iv.   Axis Bank (AXISBANK): 10.01
    v.    State Bank of India (SBIN): 9.85
    vi.   IndusInd Bank (INDUSINDBK): 5.91
    vii.  Bank of Baroda (BANKBARODA): 2.62
    viii. AU Small Finance Bank (AUBANK): 2.49
    ix.   Federal Bank (FEDERALBANK): 2.38
    x.    IDFC First Bank (IDFCFIRSTB): 1.55

Table 6 represents the annual return and volatility for banking sector stocks for the training period from Jan 1, 2017, to Dec 31, 2021. ICICIBANK produces the highest annual return while INDUSINDBK yields the lowest return. INDUSINDBK also has the highest annual risk while HDFCBANK has the lowest.

**Table 6.** The return and risk of the banking sector stocks

| Stock | Annual Return (%) | Annual Risk (%) |
|---|---|---|
| HDFCBANK | 12.28 | 24.91 |
| ICICIBANK | 25.49 | 35.63 |
| AXISBANK | 5.80 | 38.54 |
| SBIN | 14.53 | 36.87 |
| KOTAKBANK | 16.73 | 29.28 |
| INDUSINDBK | -12.52 | 48.43 |
| BANKBARODA | -11.65 | 45.39 |
| AUBANK | 12.53 | 43.05 |
| FEDERALBNK | -4.87 | 40.64 |
| IDFCFIRSTB | -0.85 | 41.21 |

Table 7 presents the weights allocation to different stocks for the three portfolio design approaches. HDFCBANK receives the highest allocation as per the minimum risk portfolio while ICICIBANK receives the highest allocation as per the optimum risk portfolio.

**Table 7.** The portfolio weights for the banking sector stocks

| Stock | EWP | MRP | MVP/ORP |
|---|---|---|---|
| HDFCBANK | 0.1 | 0.295298 | 0.058031 |
| ICICIBANK | 0.1 | 0.101177 | 0.233981 |
| AXISBANK | 0.1 | 0.003179 | 0.103966 |
| SBIN | 0.1 | 0.070569 | 0.173993 |
| KOTAKBANK | 0.1 | 0.219924 | 0.175173 |
| INDUSINDBK | 0.1 | 0.048226 | 0.018862 |
| BANKBARODA | 0.1 | 0.039036 | 0.002539 |
| AUBANK | 0.1 | 0.095645 | 0.196626 |
| FEDERALBNK | 0.1 | 0.079754 | 0.024523 |
| IDFCFIRSTB | 0.1 | 0.047192 | 0.012306 |

Figure 3 depicts the weight allocation to the banking sector stocks done by the optimum risk portfolio.

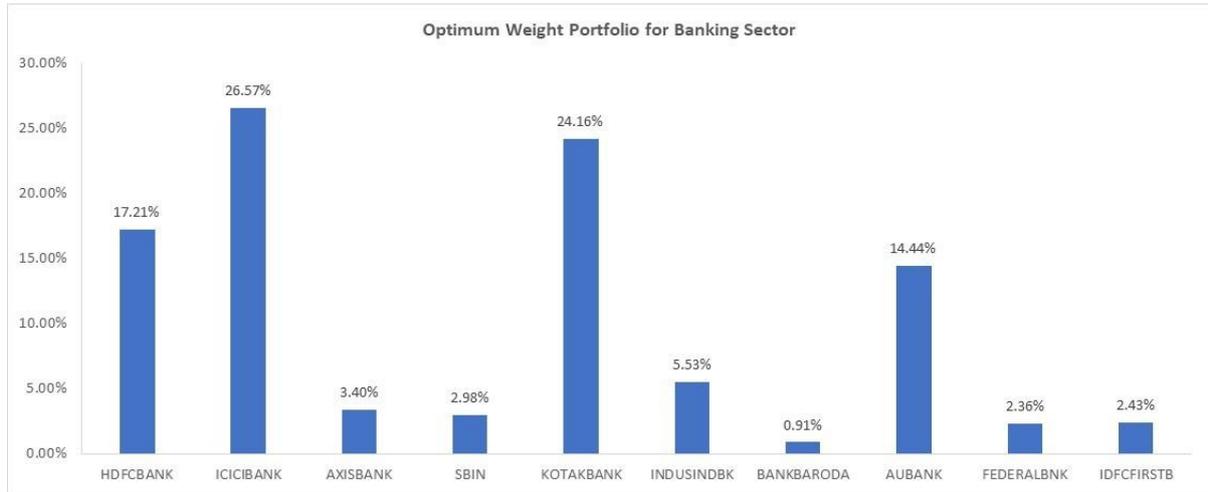

**Figure 3.** The ORP portfolio weights for the banking sector stocks

The risk and return values for the three portfolios, computed using stock prices over the training period, are depicted in Table 8. It is evident that the equal weight portfolio has yielded the lowest return and the minimum risk portfolio has exhibited the least risk. The optimum risk portfolio has produced the highest return, while the equal weight portfolio has exhibited the highest risk.

**Table 8.** The return and risk of the banking sector portfolios

| Metric | EWP | MRP | MVP (ORP) |
|---|---|---|---|
| Portfolio annual return (%) | 5.75% | 10.64% | 14.81% |
| Portfolio annual risk (%) | 27.45% | 24.21% | 26.67% |

**Table 9.** The return of the equal-weight portfolio of the banking sector stocks

| | | Date: Jan 3, 2022 | | | Date: Dec 31, 2022 | | RETURN |
|---|---|---|---|---|---|---|---|
| Stock | Weights | Price | Amount Invested | No. of Stock | Price | Value of Stock | |
| HDFCBANK | 0.1 | 1520 | 10000 | 6.58 | 1628 | 10714 | |
| ICICIBANK | 0.1 | 765 | 10000 | 13.08 | 891 | 11650 | |
| AXISBANK | 0.1 | 696 | 10000 | 14.36 | 934 | 13409 | |
| SBIN | 0.1 | 471 | 10000 | 21.24 | 614 | 13035 | |
| KOTAKBANK | 0.1 | 1824 | 10000 | 5.48 | 1827 | 10015 | 34.43% |
| INDUSINDBK | 0.1 | 912 | 10000 | 10.96 | 1220 | 13374 | |
| BANKBARODA | 0.1 | 84 | 10000 | 119.33 | 186 | 22160 | |
| AUBANK | 0.1 | 533 | 10000 | 18.77 | 654 | 12287 | |
| FEDERALBNK | 0.1 | 87 | 10000 | 114.68 | 139 | 15946 | |
| IDFCFIRSTB | 0.1 | 50 | 10000 | 201.41 | 59 | 11843 | |
| | | | 100000 | | | 134433 | |

The annual return for an investor, investing INR 100,000 on Jan 3, 2022, and following the equal weight portfolio approach is shown in Table 9. The investor receives a return of 34.43% at the end of the twelve months.

**Table 10.** The return of the optimum risk portfolio of the banking sector stocks

| Stock | Date: Jan 3, 2022 | | | | Date: Dec 31, 2022 | | RETURN |
|---|---|---|---|---|---|---|---|
| | Weights | Price | Amount Invested | No. of Stock | Price | Value of Stock | |
| HDFCBANK | 0.058031 | 1520 | 5803 | 3.82 | 1628 | 6217 | 20.25% |
| ICICIBANK | 0.233981 | 765 | 23398 | 30.6 | 891 | 27258 | |
| AXISBANK | 0.103966 | 696 | 10397 | 14.93 | 934 | 13941 | |
| SBIN | 0.173993 | 471 | 17399 | 36.96 | 614 | 22680 | |
| KOTAKBANK | 0.175173 | 1824 | 17517 | 9.6 | 1827 | 17544 | |
| INDUSINDBK | 0.018862 | 912 | 1886 | 2.07 | 1220 | 2523 | |
| BANKBARODA | 0.002539 | 84 | 254 | 3.03 | 186 | 563 | |
| AUBANK | 0.196626 | 533 | 19663 | 36.91 | 654 | 24159 | |
| FEDERALBNK | 0.024523 | 87 | 2452 | 28.12 | 139 | 3910 | |
| IDFCFIRSTB | 0.012306 | 50 | 1231 | 24.79 | 59 | 1457 | |
| | | | 100000 | | | 120252 | |

The performance of the optimum risk portfolio for the banking-sector stocks is shown in Table 10. To compare the performance of this portfolio with that of the equal-weight portfolio, the initial amount of investment of INR 100,000 is kept constant. The return yielded by the optimum risk portfolio is found to be 20.25% as depicted in Table 10.

Finally, the efficient frontier for the banking sector portfolios is presented in Figure 4. In Figure 4, the minimum risk portfolio is denoted by the red star, and the optimum risk portfolio is denoted by the green star. The stock prices from Jan 1, 2017, to Dec 31, 2021, are considered in plotting the efficient frontier of the banking sector. It is to be noted that in Figure 4, the x-axis denotes the risk while the y-axis denotes the return.

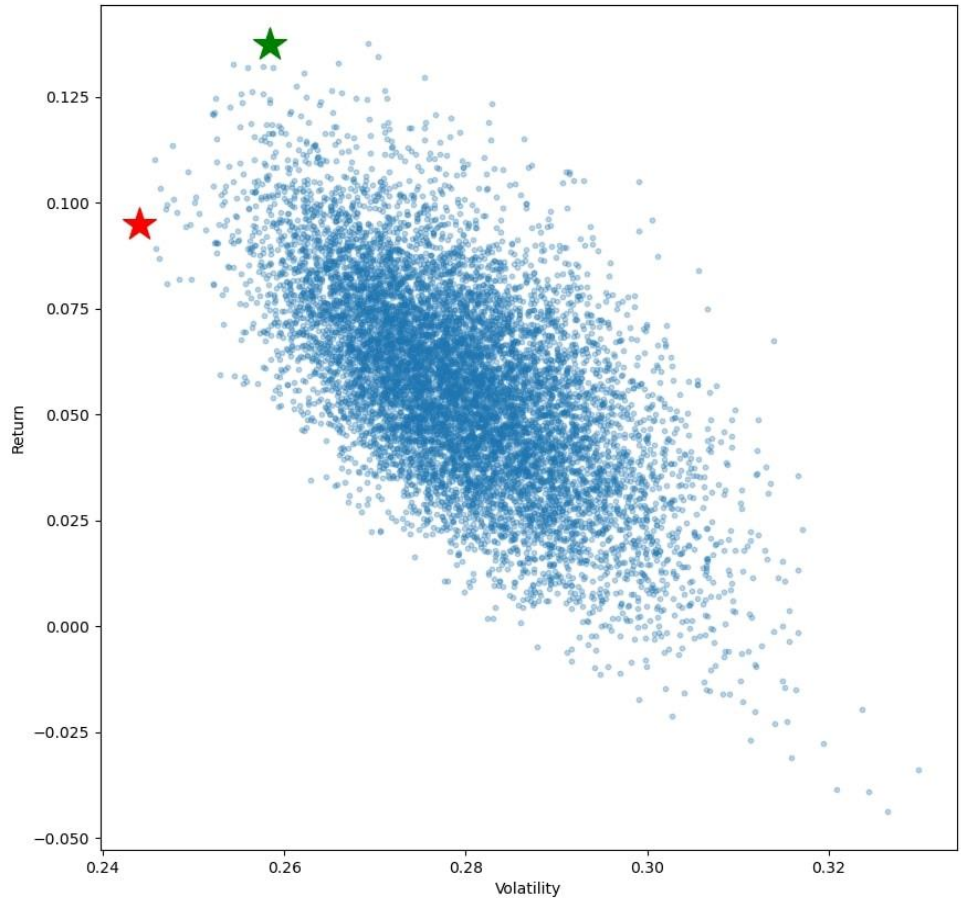

**Figure 4.** The efficient frontier of the banking sector portfolios

### 4.3 The consumer durables sector portfolios

Following are the top ten stocks of this sector based on their free-float market capitalization in the NSE, as per the report released on Dec 30, 2022. The figures below represent the contributions in percentage, of the stocks (computed based on the market capitalization) to the overall index of the consumer durables sector. The ticker name is also mentioned with a pair of parentheses just beside the name of the respective stock.

- i. Titan Company (TITAN): 32.31
- ii. Havells India (HAVELLS): 14.31
- iii. Crompton Greaves Consumer Electricals (CROMPTON): 9.87
- iv. Voltas (VOLTAS): 8.99
- v. Rajesh Exports (RAJESHEXPO): 5.61
- vi. Dixon Technologies (DIXON): 4.79
- vii. Bata India (BATAINDIA): 4.75
- viii. Kajaria Ceramics (KAJARIACER): 4.30
- ix. Blue Star (BLUESTARCO): 3.43
- x. Relaxo Footwears (RELAXO): 2.86

Table 11 represents the annual return and volatility for the consumer durables sector stocks for the training period from Jan 1, 2017, to Dec 31, 2021. DIXON shows the highest annual return and the highest risk, while RAJESHEXPO yields the lowest return and the lowest risk.

**Table 11.** The return and risk of the consumer durables sector stocks

| Stock | Annual Return (%) | Annual Risk (%) |
|---|---|---|
| TITAN | 32.24 | 33.44 |
| HAVELLS | 27.65 | 31.49 |
| CROMPTON | 15.39 | 32.40 |
| VOLTAS | 19.10 | 31.28 |
| DIXON | 98.05 | 40.36 |
| BATAINDIA | 28.72 | 29.30 |
| RAJESHEXPO | 2.97 | 27.50 |
| KAJARIACER | 22.86 | 32.44 |
| BLUESTARCO | 8.18 | 31.61 |
| RELAXO | 42.47 | 29.45 |

Table 12 presents the weights allocation to different stocks for the three portfolio design approaches. RAJESHEXPO receives the highest allocation as per the minimum risk portfolio while DIXON receives the highest allocation as per the optimum risk portfolio.

**Table 12.** The portfolio weights for the consumer durables sector stocks

| Stock | EWP | MRP | MVP/ORP |
|---|---|---|---|
| TITAN | 0.1 | 0.110719 | 0.067351 |
| HAVELLS | 0.1 | 0.069098 | 0.06275 |
| CROMPTON | 0.1 | 0.121852 | 0.033558 |
| VOLTAS | 0.1 | 0.011657 | 0.031573 |
| DIXON | 0.1 | 0.023866 | 0.329759 |
| BATAINDIA | 0.1 | 0.036811 | 0.151651 |
| RAJESHEXPO | 0.1 | 0.369238 | 0.130582 |
| KAJARIACER | 0.1 | 0.010017 | 0.012794 |
| BLUESTARCO | 0.1 | 0.100789 | 0.026107 |
| RELAXO | 0.1 | 0.145953 | 0.153873 |

Figure 5 depicts the weight allocation to the consumer durables sector stocks done by the optimum risk portfolio.

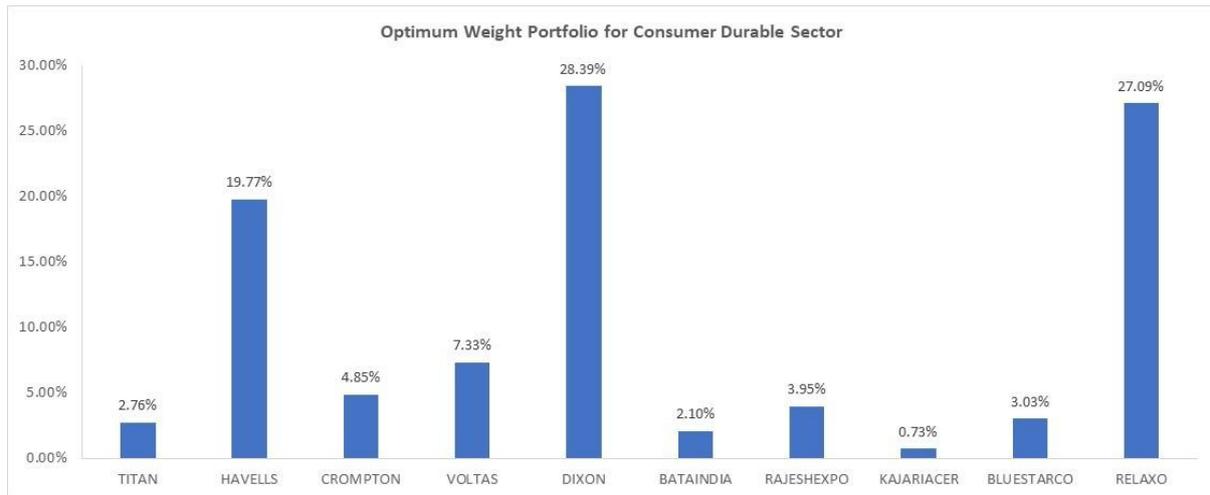

**Figure 5.** The ORP portfolio weights for the consumer durables sector stocks

The risk and return values for the three portfolios, computed using stock prices over the training period, are depicted in Table 13. It is evident that the optimum risk portfolio yields the highest return and it also exhibits the highest risk.

**Table 13.** The return and risk of the consumer durables sector portfolios

| Metric | EWP | MRP | MVP/ORP |
|---|---|---|---|
| Portfolio annual return (%) | 29.76 | 19.32 | 49.14 |
| Portfolio annual risk (%) | 18.05 | 17.01 | 20.79 |

**Table 14.** The return of the equal-weight portfolio of the consumer durables sector stocks

| Stock | Weights | Date: Jan 3, 2022 | | | Date: Dec 31, 2022 | | RETURN |
| | | Price | Amount Invested | No. of Stock | Price | Value of Stock | |
|---|---|---|---|---|---|---|---|
| TITAN | 0.1 | 2524 | 10000 | 3.96 | 2598 | 10286 | |
| HAVELLS | 0.1 | 1406 | 10000 | 7.11 | 1100 | 7821 | |
| CROMPTON | 0.1 | 441 | 10000 | 22.65 | 336 | 7621 | |
| VOLTAS | 0.1 | 1233 | 10000 | 8.11 | 800 | 6487 | |
| DIXON | 0.1 | 5517 | 10000 | 1.81 | 3905 | 7067 | -15.74% |
| BATAINDIA | 0.1 | 1859 | 10000 | 5.38 | 1649 | 8872 | |
| RAJESHEXPO | 0.1 | 853 | 10000 | 11.73 | 732 | 8583 | |
| KAJARIACER | 0.1 | 1315 | 10000 | 7.6 | 1160 | 8818 | |
| BLUESTARCO | 0.1 | 1015 | 10000 | 9.85 | 1200 | 11816 | |
| RELAXO | 0.1 | 1321 | 10000 | 7.57 | 910 | 6889 | |
| | | | 100000 | | | 84260 | |

The return of the equal-weight portfolio of the consumer durables sector stocks for an investor, investing INR 100,000 on Jan 3, 2022, is shown in Table 14. The investor receives a return of -15.74% at the end of the twelve months. The negative sign indicates that the investor has suffered a loss when using the equal-weight portfolio.

**Table 15.** The return of the optimum risk portfolio of the consumer durables sector stocks

| Stock | Date: Jan 3, 2022 | | | | Date: Dec 31, 2022 | | RETURN |
|---|---|---|---|---|---|---|---|
| | Weights | Price | Amount Invested | No. of Stock | Price | Value of Stock | |
| TITAN | 0.067351 | 2524 | 6735 | 2.67 | 2598 | 6932 | |
| HAVELLS | 0.06275 | 1406 | 6275 | 4.46 | 1100 | 4909 | |
| CROMPTON | 0.033558 | 441 | 3356 | 7.6 | 336 | 2558 | |
| VOLTAS | 0.031573 | 1233 | 3157 | 2.56 | 800 | 2048 | |
| DIXON | 0.329759 | 5517 | 32976 | 5.98 | 3905 | 23338 | -20.74% |
| BATAINDIA | 0.151651 | 1859 | 15165 | 8.16 | 1649 | 13452 | |
| RAJESHEXPO | 0.130582 | 853 | 13058 | 15.32 | 732 | 11207 | |
| KAJARIACER | 0.012794 | 1315 | 1280 | 0.97 | 1160 | 1129 | |
| BLUESTARCO | 0.026107 | 1015 | 2611 | 2.57 | 1200 | 3084 | |
| RELAXO | 0.153873 | 1321 | 15387 | 11.65 | 910 | 10599 | |
| | | | 100000 | | | 79256 | |

The performance of the optimum risk portfolio for the consumer durables sector stocks is shown in Table 15. To compare the performance of this portfolio with that of the equal-weight portfolio, the initial amount of investment of INR 100,000 is kept constant. The return yielded by the optimum risk portfolio is found to be -20.74% indicating a loss for the investor.

Finally, the efficient frontier for the consumer durables sector portfolios is presented in Figure 6. In Figure 6, the minimum risk portfolio is denoted by the red star, and the optimum risk portfolio is denoted by the green star. The stock prices from Jan 1, 2017, to Dec 31, 2021, are considered in plotting the efficient frontier of the consumer durables sector. It is to be noted that in Figure 6, the x-axis denotes the risk while the y-axis denotes the return.

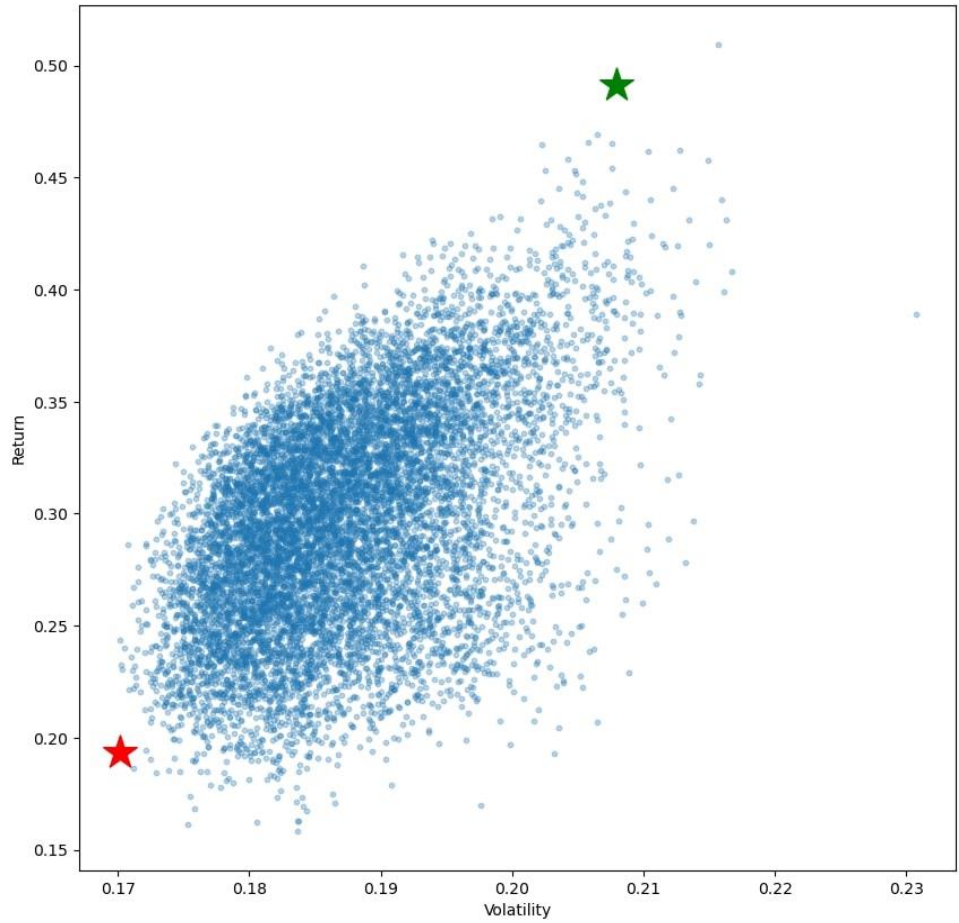

**Figure 6.** The efficient frontier of the consumer durables sector portfolios

### 4.4 The FMCG sector portfolios

Following are the top ten stocks of this sector based on their free-float market capitalization in the NSE, as per the report released on Dec 30, 2022. The figures below represent the contributions in percentage, of the stocks (computed based on the market capitalization) to the overall index of the FMCG sector. The ticker name is also mentioned with a pair of parentheses just beside the name of the respective stock.

- i. ITC (ITC): 32.39
- ii. Hindustan Unilever (HINDUNILVR): 24.00
- iii. Nestle India (NESTLEIND): 7.08
- iv. Britannia Industries (BRITANNIA): 6.23
- v. Tata Consumer Products (TATACONSUM): 5.39
- vi. Godrej Consumer Products (GODREJCP): 4.23
- vii. Dabur India (DABUR): 3.87
- viii. Varun Beverages (VBL): 3.28
- ix. Marico (MARICO): 3.15
- x. United Spirits (MCDOWELL-N): 2.80

Table 16 represents the annual return and volatility for the FMCG sector stocks for the training period from Jan 1, 2017, to Dec 31, 2021. VBL produces the highest annual return while ITC yields the lowest return. VBL also has the highest annual risk while HINDUNILVR has the lowest.

**Table 16.** The return and risk of the FMCG sector stocks

| Stock | Annual Return (%) | Annual Risk (%) |
|---|---|---|
| ITC | -4.09 | 27.58 |
| HINDUNILVR | 15.45 | 23.05 |
| NESTLEIND | 26.44 | 23.76 |
| BRITANNIA | 12.11 | 24.73 |
| TATACONSUM | 31.35 | 33.79 |
| DABUR | 13.68 | 23.57 |
| GODREJCP | 11.27 | 30.68 |
| VBL | 32.58 | 34.96 |
| MCDOWELL-N | 8.18 | 34.06 |
| MARICO | 13.11 | 24.07 |

Table 17 presents the weights allocation to different stocks for the three portfolio design approaches. ITC receives the highest allocation as per the minimum risk portfolio while TATACONSUM receives the highest allocation as per the optimum risk portfolio.

**Table 17.** The portfolio weights for the FMCG sector stocks

| Stock | EWP | MRP | MVP/ORP |
|---|---|---|---|
| ITC | 0.1 | 0.164083 | 0.02965 |
| HINDUNILVR | 0.1 | 0.103406 | 0.059826 |
| NESTLEIND | 0.1 | 0.153909 | 0.214773 |
| BRITANNIA | 0.1 | 0.132328 | 0.043768 |
| TATACONSUM | 0.1 | 0.001555 | 0.228231 |
| DABUR | 0.1 | 0.102624 | 0.016151 |
| GODREJCP | 0.1 | 0.069028 | 0.032177 |
| VBL | 0.1 | 0.121421 | 0.219175 |
| MCDOWELL-N | 0.1 | 0.012792 | 0.064606 |
| MARICO | 0.1 | 0.138854 | 0.091642 |

Figure 7 depicts the weight allocation to the FMCG sector stocks done by the optimum risk portfolio.

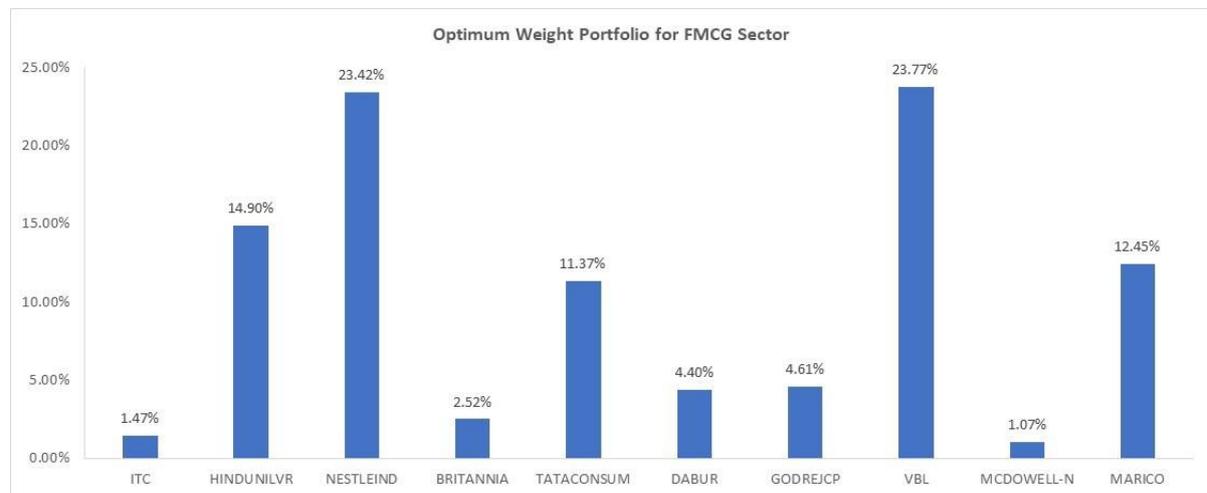

**Figure 6.** The ORP portfolio weights for the FMCG sector stocks

The risk and return values for the three portfolios, computed using stock prices over the training period, are depicted in Table 18. It is evident that the optimum risk portfolio yields the highest return and it also exhibits the highest risk.

**Table 18.** The return and risk of the FMCG sector portfolios

| Metric | EWP | MRP | MVP/ORP |
|---|---|---|---|
| Portfolio annual return (%) | 16.01 | 14.71 | 23.62 |
| Portfolio annual risk (%) | 16.57 | 15.84 | 18.13 |

**Table 19.** The return of the equal-weight portfolio of the FMCG sector stocks

| Stock | Date: Jan 3, 2022 | | | | Date: Dec 31, 2022 | | RETURN |
|---|---|---|---|---|---|---|---|
| | Weights | Price | Amount Invested | No. of Stock | Price | Value of Stock | |
| ITC | 0.1 | 219 | 10000 | 45.64 | 332 | 15132 | |
| HINDUNILVR | 0.1 | 2361 | 10000 | 4.23 | 2561 | 10846 | |
| NESTLEIND | 0.1 | 19678 | 10000 | 0.51 | 19606 | 9963 | |
| BRITANNIA | 0.1 | 3618 | 10000 | 2.76 | 4307 | 11907 | 19.10% |
| TATACONSUM | 0.1 | 748 | 10000 | 13.37 | 767 | 10253 | |
| DABUR | 0.1 | 581 | 10000 | 17.21 | 561 | 9661 | |
| GODREJCP | 0.1 | 958 | 10000 | 10.44 | 874 | 9128 | |
| VBL | 0.1 | 587 | 10000 | 17.05 | 1323 | 22549 | |
| MCDOWELL-N | 0.1 | 901 | 10000 | 11.1 | 878 | 9744 | |
| MARICO | 0.1 | 514 | 10000 | 19.45 | 510 | 9917 | |
| | | | 100000 | | | 119100 | |

The annual return for an investor, investing INR 100,000 on Jan 3, 2022, and following the equal weight portfolio approach is shown in Table 19. The investor receives a return of 19.10% at the end of the twelve months.

**Table 20.** The return of the optimum risk portfolio of the FMCG sector stocks

| Stock | Date: Jan 3, 2022 | | | | Date: Dec 31, 2022 | | RETURN |
|---|---|---|---|---|---|---|---|
| | Weights | Price | Amount Invested | No. of Stock | Price | Value of Stock | |
| ITC | 0.02965 | 219 | 2965 | 13.53 | 332 | 4487 | |
| HINDUNILVR | 0.059826 | 2361 | 5983 | 2.53 | 2561 | 6489 | |
| NESTLEIND | 0.214773 | 19678 | 21477 | 1.09 | 19606 | 21399 | |
| BRITANNIA | 0.043768 | 3618 | 4377 | 1.21 | 4307 | 5211 | 30.29% |
| TATACONSUM | 0.228231 | 748 | 22823 | 30.51 | 767 | 23401 | |
| DABUR | 0.016151 | 581 | 1615 | 2.78 | 561 | 1560 | |
| GODREJCP | 0.032177 | 958 | 3218 | auto3.36 | 874 | 2937 | |
| VBL | 0.219175 | 587 | 21918 | 37.37 | 1323 | 49421 | |
| MCDOWELL-N | 0.064606 | 901 | 6460 | 7.17 | 878 | 6295 | |
| MARICO | 0.091642 | 514 | 9164 | 17.83 | 510 | 9088 | |
| | | | 100000 | | | 130288 | |

The performance of the optimum risk portfolio for the FMCG sector stocks is shown in Table 20. To compare the performance of this portfolio with that of the equal-weight portfolio, the initial amount of investment of INR 100,000 is kept constant. The return yielded by the optimum risk portfolio is found to be 30.29% as depicted in Table 20.

Finally, the efficient frontier for the FMCG sector portfolios is presented in Figure 8. In Figure 8, the minimum risk portfolio is denoted by the red star, and the optimum risk portfolio is denoted by the green star. The stock prices from Jan 1, 2017, to Dec 31, 2021, are considered in plotting the efficient frontier of the FMCG sector. It is to be noted that in Figure 8, the x-axis denotes the risk while the y-Axis denotes the return.

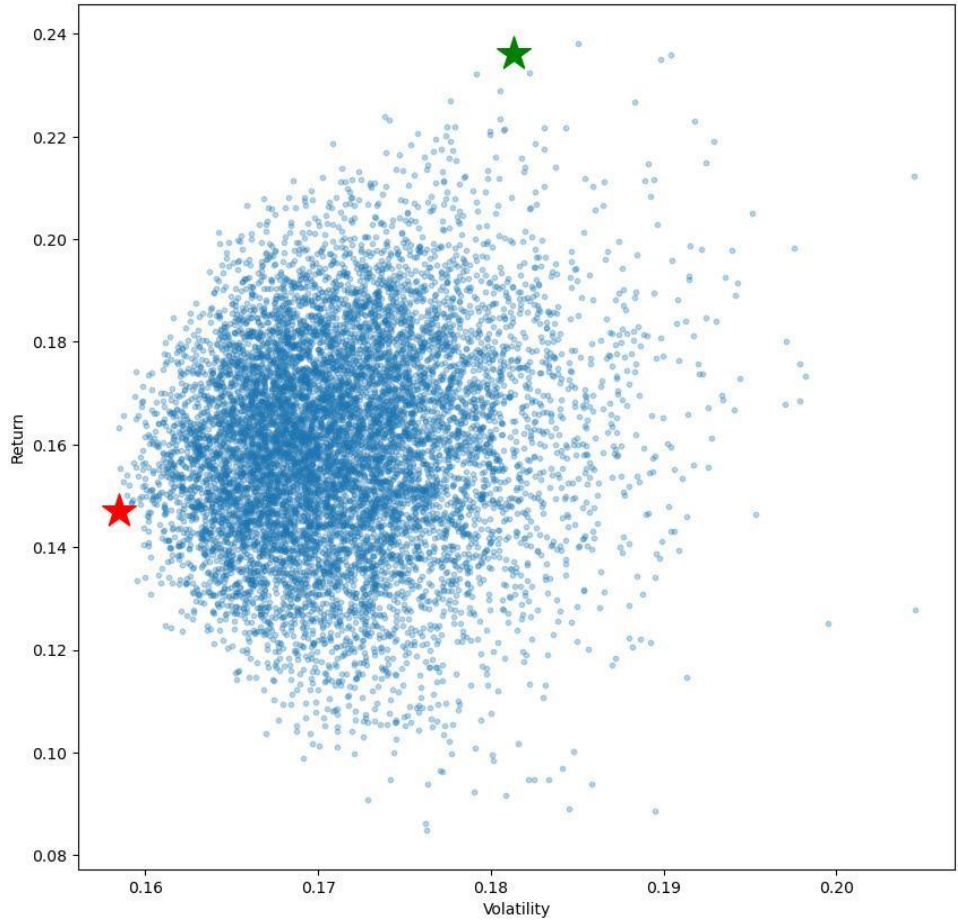

**Figure 8.** The efficient frontier of the FMCG sector portfolio

### 4.5 The financial services sector portfolios

Following are the top ten stocks of this sector based on their free-float market capitalization in the NSE, as per the report released on Dec 30, 2022. The figures below represent the contributions in percentage, of the stocks (computed based on the market capitalization) to the overall index of the financial services sector. The ticker name is also mentioned with a pair of parentheses just beside the name of the respective stock.

    i.      HDFC Bank (HDFCBANK): 23.62
    ii.     ICICI Bank (ICICIBANK): 19.41
    iii.    Housing Development Finance Corporation (HDFC): 15.82
    iv.    Kotak Mahindra Bank (KOTAKBANK): 8.39
    v.     Axis Bank (AXISBANK): 7.89
    vi.    State Bank of India (SBIN): 7.10
    vii.   Bajaj Finance (BAJFINANCE): 5.24

viii. Bajaj Finserv (BAJAJFINSV): 2.43
ix. HDFC Life Insurance Company (HDFCLIFE): 1.91
x. SBI Life Insurance Company (SBILIFE): 1.84

Table 21 represents the annual return and volatility for the financial services sector stocks for the training period from Jan 1, 2017, to Dec 31, 2021. BAJAJFINANCE shows the highest annual return while AXISBANK yields the lowest. BAJAJFINANCE also has the highest annual risk while HDFCBANK has the lowest.

**Table 21.** The return and risk of the financial services sector stocks

| Stock | Annual Return (%) | Annual Risk (%) |
|---|---|---|
| HDFCBANK | 12.28 | 24.91 |
| ICICIBANK | 25.49 | 35.63 |
| HDFC | 11.20 | 30.27 |
| KOTAKBANK | 16.73 | 29.28 |
| AXISBANK | 5.80 | 38.54 |
| SBIN | 14.53 | 36.87 |
| BAJFINANCE | 41.87 | 40.69 |
| BAJAJFINSV | 36.97 | 37.23 |
| SBILIFE | 18.28 | 30.20 |
| HDFCLIFE | 16.52 | 32.64 |

Table 22 presents the weights allocation to different stocks for the three portfolio design approaches. SBILIFE receives the highest allocation as per the minimum risk portfolio while BAJAJFINANCE receives the highest allocation as per the optimum risk portfolio.

**Table 22.** The portfolio weights for the financial services sector stocks

| Stock | EWP | MRP | MVP/ORP |
|---|---|---|---|
| HDFCBANK | 0.1 | 0.226283 | 0.091791 |
| ICICIBANK | 0.1 | 0.000271 | 0.100556 |
| HDFC | 0.1 | 0.134293 | 0.001019 |
| KOTAKBANK | 0.1 | 0.212987 | 0.048596 |
| AXISBANK | 0.1 | 0.019966 | 0.029816 |
| SBIN | 0.1 | 0.025269 | 0.011026 |
| BAJFINANCE | 0.1 | 0.002448 | 0.243249 |
| BAJAJFINSV | 0.1 | 0.055421 | 0.215871 |
| SBILIFE | 0.1 | 0.235425 | 0.190135 |
| HDFCLIFE | 0.1 | 0.087638 | 0.067941 |

Figure 9 depicts the weight allocation to the financial services sector stocks done by the optimum risk portfolio.

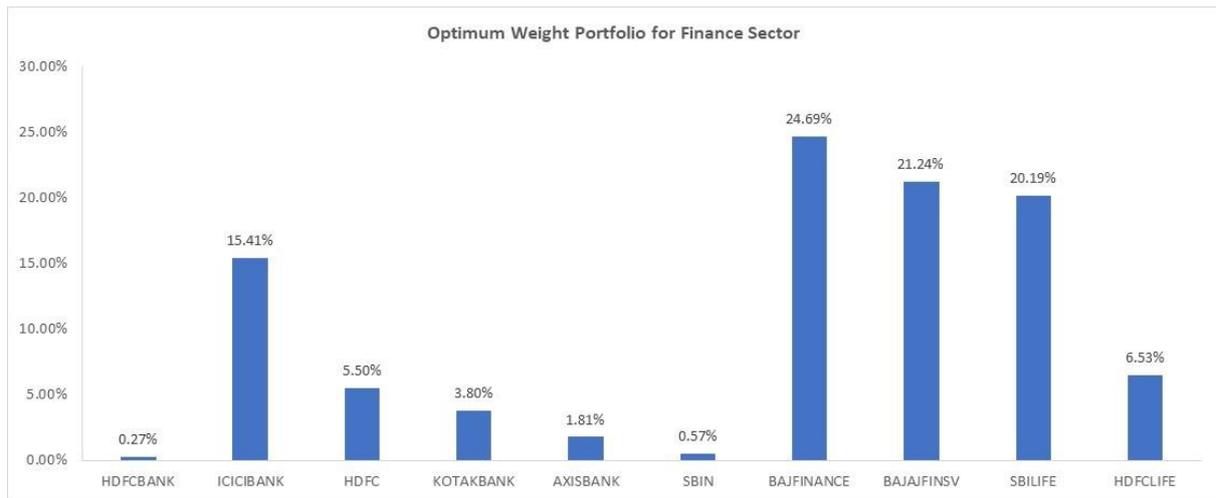

**Figure 7.** The ORP portfolio weights for the financial services sector stocks

The risk and return values for the three portfolios, computed using stock prices over the training period, are depicted in Table 23. It is evident that the optimum risk portfolio yields the highest return and it also exhibits the highest risk.

**Table 23.** The return and risk of the financial services sector portfolios

| Metric | EWP | MRP | MVP/ORP |
|---|---|---|---|
| Portfolio annual return (%) | 19.97 | 16.24 | 27.61 |
| Portfolio annual risk (%) | 24.78 | 22.23 | 26.91 |

**Table 24.** The return of the equal-weight portfolio of the financial services sector stocks

| Stock | Date: Jan 3, 2022 | | | | Date: Dec 31, 2022 | | RETURN |
|---|---|---|---|---|---|---|---|
| | Weights | Price | Amount Invested | No. of Stock | Price | Value of Stock | |
| HDFCBANK | 0.1 | 1520 | 10000 | 6.58 | 1628 | 10714 | |
| ICICIBANK | 0.1 | 765 | 10000 | 13.08 | 891 | 11650 | |
| HDFC | 0.1 | 2636 | 10000 | 3.79 | 2638 | 10005 | |
| KOTAKBANK | 0.1 | 1824 | 10000 | 5.48 | 1827 | 10015 | 5.94% |
| AXISBANK | 0.1 | 696 | 10000 | 14.36 | 934 | 13409 | |
| SBIN | 0.1 | 471 | 10000 | 21.24 | 614 | 13035 | |
| BAJFINANCE | 0.1 | 7220 | 10000 | 1.39 | 6575 | 9107 | |
| BAJAJFINSV | 0.1 | 1698 | 10000 | 5.89 | 1548 | 9115 | |
| SBILIFE | 0.1 | 1209 | 10000 | 8.27 | 1231 | 10181 | |
| HDFCLIFE | 0.1 | 651 | 10000 | 15.37 | 566 | 8705 | |
| | | | 100000 | | | 105936 | |

The return of the equal-weight portfolio of the financial services sector stocks for an investor, investing INR 100,000 on Jan 3, 2022, is shown in Table 24. The investor receives a return of 5.94% at the end of the twelve months.

**Table 25.** The return of the optimum risk portfolio of the financial services sector stocks

| Stock | Date: Jan 3, 2022 | | | | Date: Dec 31, 2022 | | RETURN |
|---|---|---|---|---|---|---|---|
| | Weights | Price | Amount Invested | No. of Stock | Price | Value of Stock | |
| HDFCBANK | 0.091791 | 1520 | 9179 | 6.04 | 1628 | 9834 | |
| ICICIBANK | 0.100556 | 765 | 10056 | 13.15 | 891 | 11714 | |
| HDFC | 0.001019 | 2636 | 101 | 0.04 | 2638 | 102 | |
| KOTAKBANK | 0.048596 | 1824 | 4860 | 2.66 | 1827 | 4867 | -0.94% |
| AXISBANK | 0.029816 | 696 | 2981 | 4.28 | 934 | 3998 | |
| SBIN | 0.011026 | 471 | 1103 | 2.34 | 614 | 1437 | |
| BAJFINANCE | 0.243249 | 7220 | 24325 | 3.37 | 6575 | 22154 | |
| BAJAJFINSV | 0.215871 | 1698 | 21587 | 12.71 | 1548 | 19677 | |
| SBILIFE | 0.190135 | 1209 | 19014 | 15.72 | 1231 | 19358 | |
| HDFCLIFE | 0.067941 | 651 | 6794 | 10.44 | 566 | 5914 | |
| | | | 100000 | | | 99055 | |

The performance of the optimum risk portfolio for the financial services sector stocks is shown in Table 25. To compare the performance of this portfolio with that of the equal-weight portfolio, the initial amount of investment of INR 100,000 is kept constant. The return yielded by the optimum risk portfolio is found to be -0.94% indicating a loss for the investor. The negative sign indicates that the investor has suffered a loss.

Finally, the efficient frontier for the financial services sector portfolios is presented in Figure 10. In Figure 10, the minimum risk portfolio is denoted by the red star, and the optimum risk portfolio is denoted by the green star. The stock prices from Jan 1, 2017, to Dec 31, 2021, are considered in plotting the efficient frontier of the financial services sector. It is to be noted that in Figure 10, the x-axis denotes the risk while the y-axis denotes the return.

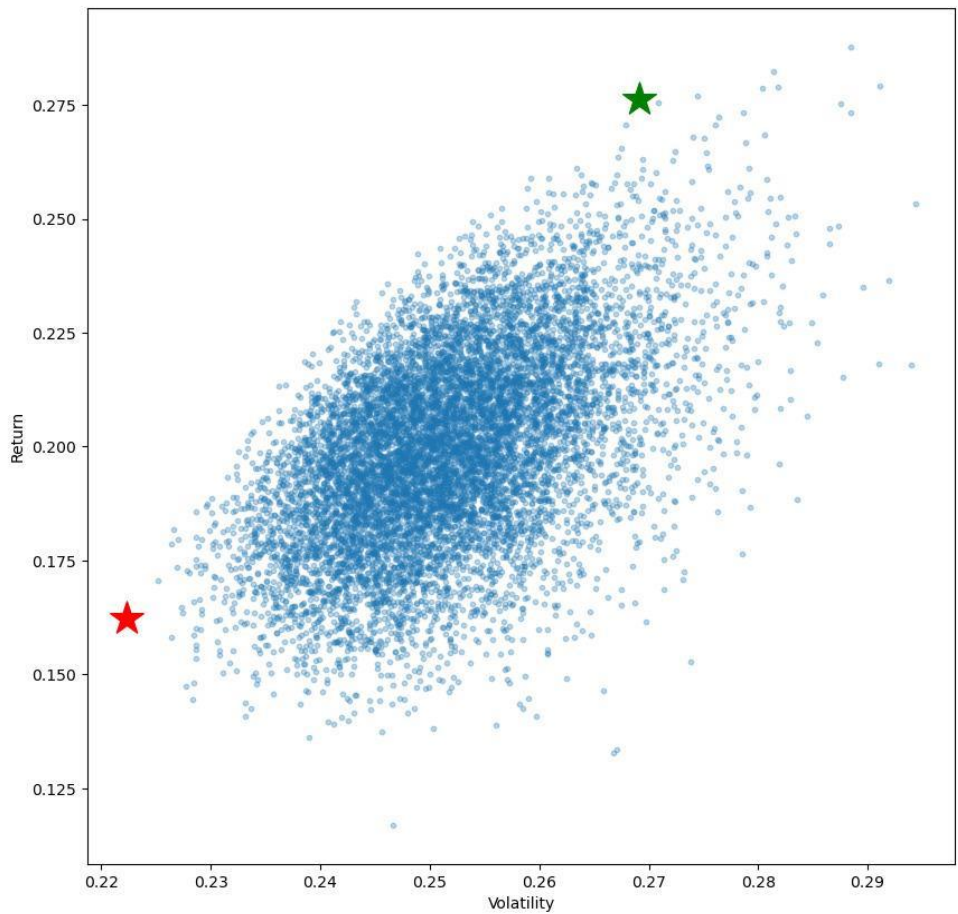

**Figure 8.** The efficient frontier of the financial services sector portfolios

## 4.6 The IT sector portfolios

Following are the top ten stocks of this sector based on their free-float market capitalization in the NSE, as per the report released on Dec 30, 2022. The figures below represent the contributions in percentage, of the stocks (computed based on the market capitalization) to the overall index of the IT sector. The ticker name is also mentioned with a pair of parentheses just beside the name of the respective stock.

    i.     Tata Consultancy Services (TCS): 26.41
    ii.    Infosys (INFY): 25.95
    iii.   HCL Technologies (HCLTECH): 9.49
    iv.   Wipro (WIPRO): 9.04
    v.    Tech Mahindra (TECHM): 8.81
    vi.   LTI Mindtree (LTIM): 7.67
    vii.  Persistent Systems (PERSISTENT): 4.52
    viii. MphasiS (MPHASIS): 3.29
    ix.   Coforge (COFORGE): 3.07
    x.    L&T Technology Services (LTTS): 1.75

Table 26 represents the annual return and volatility for the IT sector stocks for the training period from Jan 1, 2017, to Dec 31, 2021. PERSISTENT shows the highest annual return while TCS yields the lowest. COFORGE has the highest annual risk while TCS has the lowest.

**Table 26.** The return and risk of the IT sector stocks

| Stock | Annual Return (%) | Annual Risk (%) |
|---|---|---|
| INFY | 39.88 | 28.05 |
| TCS | 29.34 | 25.48 |
| HCLTECH | 33.02 | 27.81 |
| TECHM | 40.13 | 31.08 |
| WIPRO | 36.67 | 26.87 |
| LTIM | 66.31 | 34.90 |
| PERSISTENT | 85.82 | 33.93 |
| MPHASIS | 54.61 | 34.10 |
| COFORGE | 76.10 | 43.83 |
| LTTS | 62.68 | 38.15 |

Table 27 presents the weights allocation to different stocks for the three portfolio design approaches. TCS receives the highest allocation as per the minimum risk portfolio while PERSISTENT receives the highest allocation as per the optimum risk portfolio.

**Table 27.** The portfolio weights for the IT sector stocks

| Stock | EWP | MRP | MVP/ORP |
|---|---|---|---|
| INFY | 0.1 | 0.11313 | 0.085182 |
| TCS | 0.1 | 0.210903 | 0.004158 |
| HCLTECH | 0.1 | 0.096888 | 0.085747 |
| TECHM | 0.1 | 0.035844 | 0.054446 |
| WIPRO | 0.1 | 0.203684 | 0.024097 |
| LTIM | 0.1 | 0.07319 | 0.141618 |
| PERSISTENT | 0.1 | 0.171321 | 0.294376 |
| MPHASIS | 0.1 | 0.070874 | 0.150793 |
| COFORGE | 0.1 | 0.002889 | 0.035054 |
| LTTS | 0.1 | 0.021276 | 0.124528 |

Figure 11 depicts the weight allocation to the IT sector stocks done by the optimum risk portfolio.

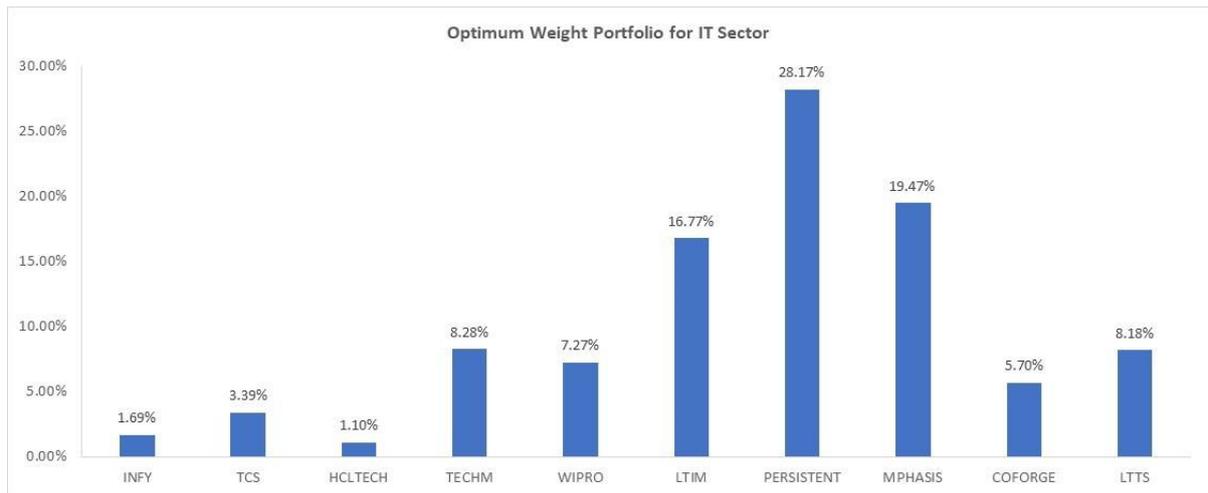

**Figure 9.** The ORP portfolio weights for the financial services sector stocks

The risk and return values for the three portfolios, computed using stock prices over the training period, are depicted in Table 28. It is evident that the optimum risk portfolio yields the highest return and it also exhibits the highest risk.

**Table 28.** The return and risk of the IT sector portfolios

| Metric | EWP | MRP | MVP/ORP |
|---|---|---|---|
| Portfolio annual return (%) | 52.46 | 47.79 | 62.78 |
| Portfolio annual risk (%) | 21.14 | 19.71 | 22.03 |

**Table 29.** The return of the equal-weight portfolio of the IT sector stocks

| | Date: Jan 3, 2022 | | | | Date: Dec 31, 2022 | | RETURN |
|---|---|---|---|---|---|---|---|
| Stock | Weights | Price | Amount Invested | No. of Stock | Price | Value of Stock | |
| INFY | 0.1 | 1898 | 10000 | 5.27 | 1508 | 7944 | |
| TCS | 0.1 | 3818 | 10000 | 2.62 | 3257 | 8530 | |
| HCLTECH | 0.1 | 1326 | 10000 | 7.54 | 1039 | 7837 | |
| TECHM | 0.1 | 1785 | 10000 | 5.6 | 1016 | 5695 | -32.09% |
| WIPRO | 0.1 | 719 | 10000 | 13.91 | 393 | 5465 | |
| LTIM | 0.1 | 7533 | 10000 | 1.33 | 4365 | 5795 | |
| PERSISTENT | 0.1 | 4872 | 10000 | 2.05 | 3871 | 7945 | |
| MPHASIS | 0.1 | 3423 | 10000 | 2.92 | 1973 | 5764 | |
| COFORGE | 0.1 | 5973 | 10000 | 1.67 | 3884 | 6503 | |
| LTTS | 0.1 | 5727 | 10000 | 1.75 | 3684 | 6432 | |
| | | | 100000 | | | 67910 | |

The return of the equal-weight portfolio of the IT sector stocks for an investor, investing INR 100,000 on Jan 3, 2022, is shown in Table 29. The investor receives a return of -32.09% at the end of the twelve months. This indicates that the investor has suffered a massive loss using the equal-weight portfolio.

**Table 30.** The return of the optimum risk portfolio of the IT sector stocks

| Stock | Date: Jan 3, 2022 | | | | Date: Dec 31, 2022 | | RETURN |
|---|---|---|---|---|---|---|---|
| | Weights | Price | Amount Invested | No. of Stock | Price | Value of Stock | |
| INFY | 0.085182 | 1898 | 8518 | 4.49 | 1508 | 6767 | |
| TCS | 0.004158 | 3818 | 416 | 0.11 | 3257 | 355 | |
| HCLTECH | 0.085747 | 1326 | 8575 | 6.47 | 1039 | 6720 | |
| TECHM | 0.054446 | 1785 | 5445 | 3.05 | 1016 | 3101 | -31.16% |
| WIPRO | 0.024097 | 719 | 2410 | 3.35 | 393 | 1317 | |
| LTIM | 0.141618 | 7533 | 14161 | 1.88 | 4365 | 8207 | |
| PERSISTENT | 0.294376 | 4872 | 29438 | 6.04 | 3871 | 23388 | |
| MPHASIS | 0.150793 | 3423 | 15079 | 4.41 | 1973 | 8692 | |
| COFORGE | 0.035054 | 5973 | 3505 | 0.59 | 3884 | 2279 | |
| LTTS | 0.124528 | 5727 | 12453 | 2.17 | 3684 | 8010 | |
| | | | 100000 | | | 68836 | |

The performance of the optimum risk portfolio for the IT sector stocks is shown in Table 30. To compare the performance of this portfolio with that of the equal-weight portfolio, the initial amount of investment of INR 100,000 is kept constant. The return yielded by the optimum risk portfolio is found to be -31.16% indicating a loss for the investor. Again, the investor has suffered a big loss.

Finally, the efficient frontier for the IT sector portfolios is presented in Figure 12. In Figure 12, the minimum risk portfolio is denoted by the red star, and the optimum risk portfolio is denoted by the green star. The stock prices from Jan 1, 2017, to Dec 31, 2021, are considered in plotting the efficient frontier of the IT sector. It is to be noted that in Figure 12, the x-axis denotes the risk while the y-axis denotes the return.

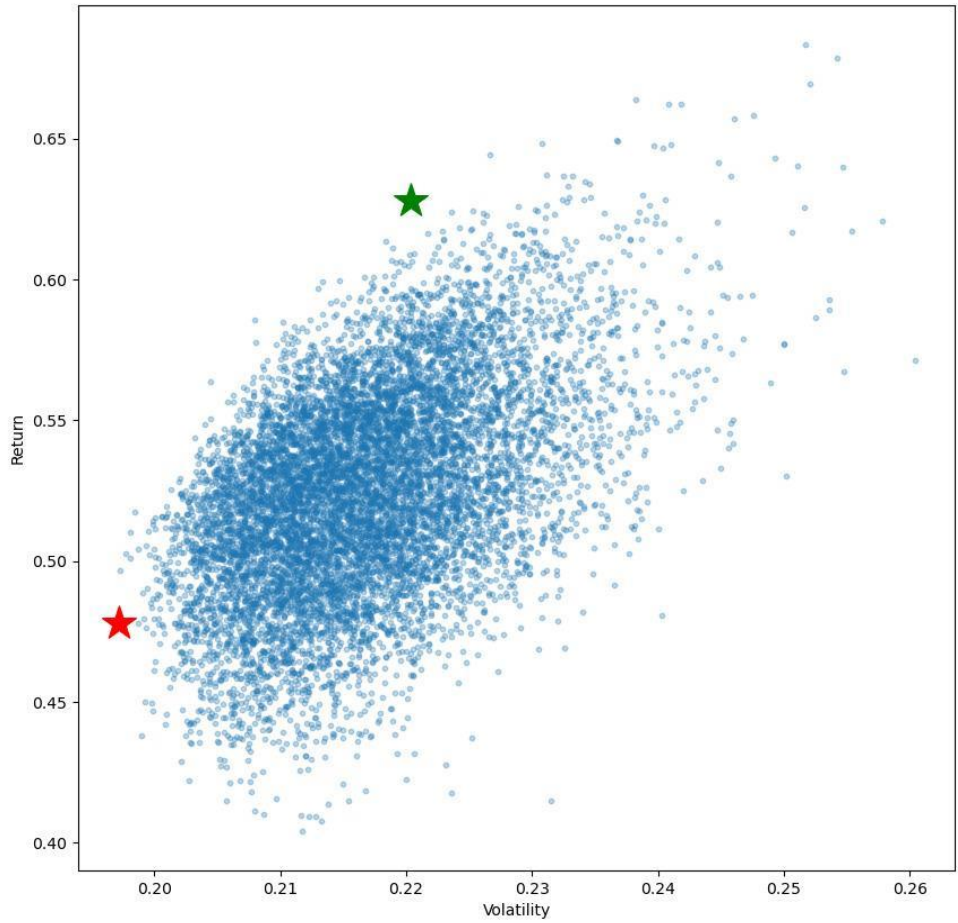

**Figure 12.** The efficient frontier of the financial services sector portfolios

### 4.7 The media sector portfolios

Following are the top ten stocks of this sector based on their free-float market capitalization in the NSE, as per the report released on Dec 30, 2022. The figures below represent the contributions in percentage, of the stocks (computed based on the market capitalization) to the overall index of the media sector. The ticker name is also mentioned with a pair of parentheses just beside the name of the respective stock.

i. Zee Entertainment Enterprise (ZEEL): 26.82
ii. PVR (PVR): 24.64
iii. TV18 Broadcast (TV18BRDCST): 9.15
iv. Sun TV Network (SUNTV): 8.91
v. Dish TV India (DISHTV): 8.07
vi. Nazara Tech (NAZARA): 7.10
vii. Network18 Media & Investments (NETWORK18) : 6.52
viii. Navneet Education (NAVNETEDU): 3.68
ix. Hathway Cable and Datacom (HATHWAY): 2.98
x. NDTV (NDTV): 2.12

The stock of NAZARA was found to contain more than 30% of its observations missing during the training period from Jan 1, 2017, to Dec 31, 2021. The first observation of this stock was available on March 30, 2021. Hence, for NAZARA, only 15% of the observations were available. The computation for the portfolios for the media sector, therefore, did not involve the stock of NAZARA. The remaining nine stocks are used in the portfolio design.

Table 31 represents the annual return and volatility for the media sector stocks for the training period from Jan 1, 2017, to Dec 31, 2021. NDTV shows the highest annual return while DISHTV yields the lowest. DISHTV also has the highest annual risk while NAVNETEDUL has the lowest.

**Table 31.** The return and risk of the media sector stocks

| Stock | Annual Return (%) | Annual Risk (%) |
|---|---|---|
| ZEEL | -9.19 | 53.91 |
| PVR | -0.02 | 39.66 |
| TV18BRDCST | 2.21 | 50.36 |
| SUNTV | -13.27 | 39.21 |
| NAVNETEDUL | -12.89 | 34.08 |
| DISHTV | -19.78 | 64.21 |
| NETWORK18 | 31.68 | 50.79 |
| NDTV | 49.24 | 51.31 |
| HATHWAY | -6.70 | 58.83 |

Table 32 presents the weights allocation to different stocks for the three portfolio design approaches. NAVNETEDUL receives the highest allocation as per the minimum risk portfolio while NETWORK18 receives the highest allocation as per the optimum risk portfolio.

**Table 32.** The portfolio weights for the media sector stocks

| Stock | EWP | MRP | MVP/ORP |
|---|---|---|---|
| ZEEL | 0.1 | 0.067996 | 0.063982 |
| PVR | 0.1 | 0.102364 | 0.057247 |
| TV18BRDCST | 0.1 | 0.031983 | 0.110087 |
| SUNTV | 0.1 | 0.152141 | 0.01789 |
| NAVNETEDUL | 0.1 | 0.365526 | 0.143365 |
| DISHTV | 0.1 | 0.002639 | 0.008657 |
| NETWORK18 | 0.1 | 0.005777 | 0.317699 |
| NDTV | 0.1 | 0.141213 | 0.266208 |
| HATHWAY | 0.1 | 0.130359 | 0.014864 |

Figure 13 depicts the weight allocation to the media sector stocks done by the optimum risk portfolio.

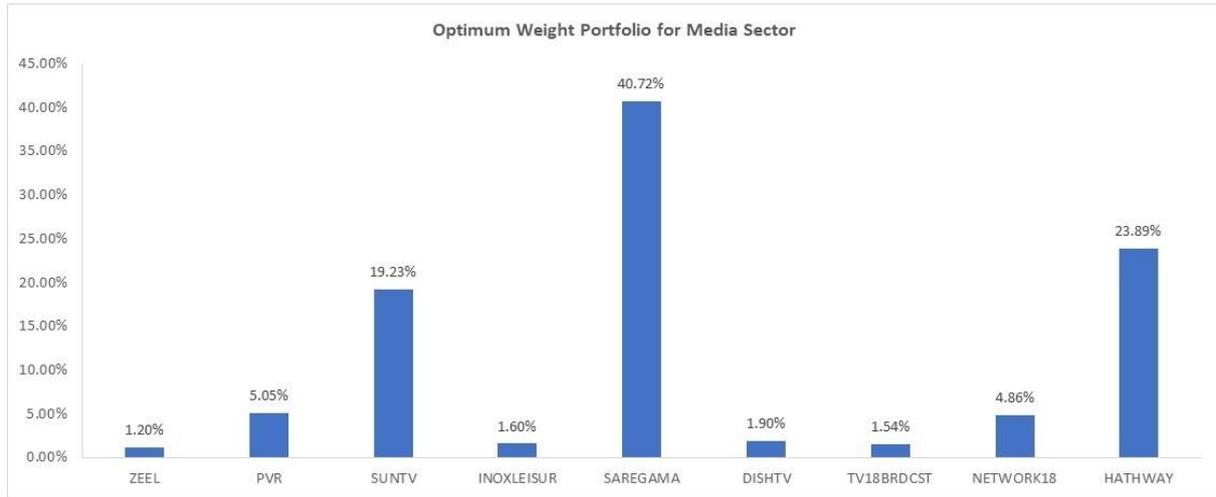

**Figure 10.** The ORP portfolio weights for the media sector stocks

The risk and return values for the three portfolios, computed using stock prices over the training period, are depicted in Table 33. It is evident that the optimum risk portfolio yields the highest return and it also exhibits the highest risk.

**Table 33.** The return and risk of the media sector portfolios

| Metric | EWP | MRP | MVP/ORP |
|---|---|---|---|
| Portfolio annual return (%) | 2.13% | -1.08% | 20.47% |
| Portfolio annual risk (%) | 24.66% | 24.44% | 30.23% |

**Table 34.** The return of the equal-weight portfolio of the media sector stocks

| | Date: Jan 3, 2022 | | | | Date: Dec 31, 2022 | | RETURN |
|---|---|---|---|---|---|---|---|
| Stock | Weights | Price | Amount Invested | No. of Stock | Price | Value of Stock | |
| ZEEL | 0.1 | 323 | 10000 | 30.96 | 240 | 7433 | |
| PVR | 0.1 | 1341 | 10000 | 7.46 | 1720 | 12827 | |
| TV18BRDCST | 0.1 | 504 | 10000 | 19.83 | 487 | 9648 | |
| SUNTV | 0.1 | 355 | 10000 | 28.13 | 500 | 14055 | -6.13% |
| NAVNETEDUL | 0.1 | 539 | 10000 | 18.56 | 385 | 7152 | |
| DISHTV | 0.1 | 18 | 10000 | 558.66 | 18 | 10279 | |
| NETWORK18 | 0.1 | 46 | 10000 | 216.22 | 37 | 8011 | |
| NDTV | 0.1 | 92 | 10000 | 108.87 | 66 | 7197 | |
| HATHWAY | 0.1 | 22 | 10000 | 455.58 | 17 | 7882 | |
| | | | 90000 | | | 84484 | |

The annual return for an investor, investing INR 100,000 on Jan 3, 2022, following the equal weight portfolio approach is shown in Table 34. The annual return in this case is -6.13%. The negative value indicates that the investor has incurred a loss.

**Table 35.** The return of the optimum risk portfolio of the media sector stocks

| Stock | Date: Jan 3, 2022 | | | | Date: Dec 31, 2022 | | RETURN |
| --- | --- | --- | --- | --- | --- | --- | --- |
| | Weights | Price | Amount Invested | No. of Stock | Price | Value of Stock | |
| ZEEL | 0.063982 | 323 | 6398 | 19.81 | 240 | 4756 | |
| PVR | 0.057247 | 1341 | 5725 | 4.27 | 1720 | 7343 | |
| TV18BRDCST | 0.110087 | 504 | 11009 | 21.83 | 487 | 10621 | |
| SUNTV | 0.01789 | 355 | 1789 | 5.03 | 500 | 2515 | -17.84% |
| NAVNETEDUL | 0.143365 | 539 | 14337 | 26.6 | 385 | 10253 | |
| DISHTV | 0.008657 | 18 | 866 | 48.36 | 18 | 890 | |
| NETWORK18 | 0.317699 | 46 | 31770 | 686.92 | 37 | 25450 | |
| NDTV | 0.266208 | 92 | 26620 | 289.83 | 66 | 19158 | |
| HATHWAY | 0.014864 | 22 | 1486 | 67.72 | 17 | 1172 | |
| | | | 100000 | | | 82158 | |

The performance of the optimum risk portfolio for the media sector stocks is shown in Table 35. To compare the performance of this portfolio with that of the equal-weight portfolio, the initial amount of investment of INR 100,000 is kept constant. The return yielded by the optimum risk portfolio is found to be -17.84%, indicating a loss for the investor.

Finally, the efficient frontier for the media sector portfolios is presented in Figure 14. In Figure 14, the minimum risk portfolio is denoted by the red star, and the optimum risk portfolio is denoted by the green star. The stock prices from Jan 1, 2017, to Dec 31, 2021, are considered in plotting the efficient frontier of the media sector. It is to be noted that in Figure 14, the x-axis denotes the risk while the y-axis denotes the return.

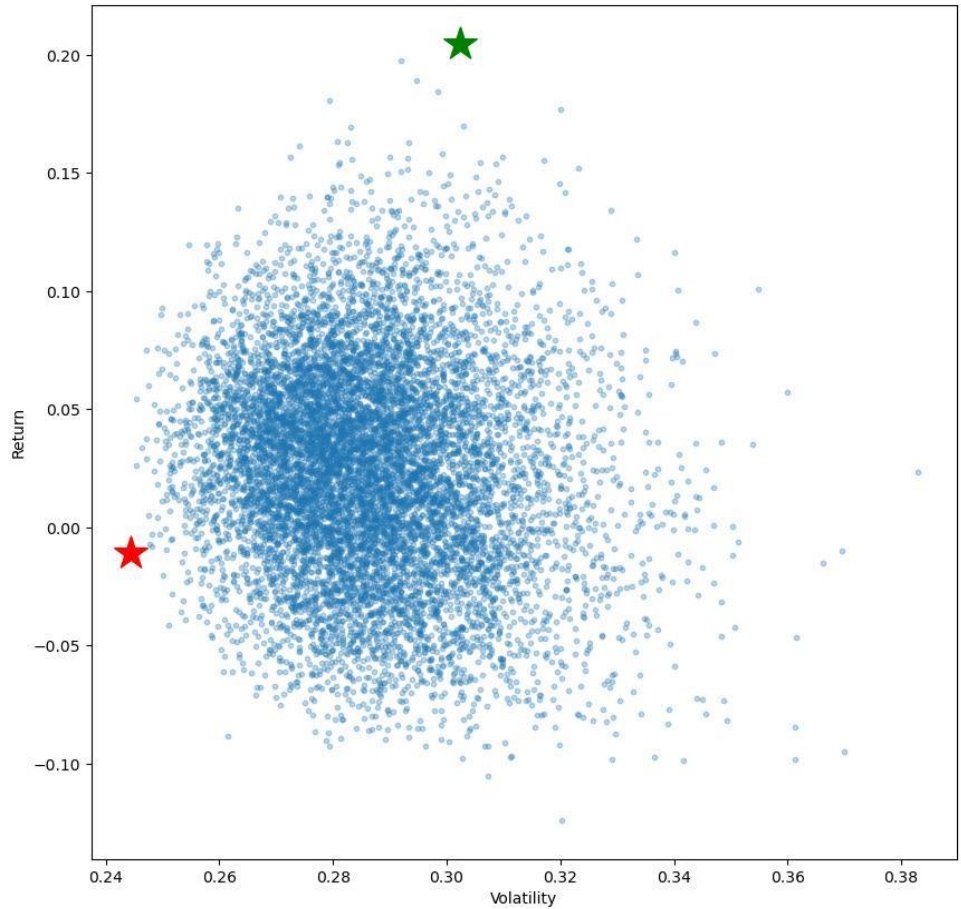

**Figure 11**. The efficient frontier of the media sector portfolios

## 4.8 The metal sector portfolios

Following are the top ten stocks of this sector based on their free-float market capitalization in the NSE, as per the report released on Dec 30, 2022. The figures below represent the contributions in percentage, of the stocks (computed based on the market capitalization) to the overall index of the metal sector. The ticker name is also mentioned with a pair of parentheses just beside the name of the respective stock.

   i. Tata Steel (TATASTEEL): 21.98
  ii. Adani Enterprises (ADANIENT): 18.52
 iii. Hindalco Industries (HINDALCO): 15.57
  iv. JSW Steel (JSWSTEEL): 15.37
   v. Vedanta (VEDL): 8.44
  vi. Jindal Steel & Power (JINDALSTEL): 5.01
 vii. APL Apollo Tubes (APLAPOLLO): 4.48
viii. Steel Authority of India (SAIL): 2.98
  ix. Hindustan Zinc (HINDZINC): 1.92
   x. National Aluminium Co. (NATIONALUM): 1.71

Table 36 represents the annual return and volatility for the metal sector stocks for the training period from Jan 1, 2017, to Dec 31, 2021. ADANIENT shows the highest annual return while HINDZINC yields the lowest. JINDALSTEL has the highest annual risk while HINDZINC has the lowest.

**Table 36.** The return and risk of metal sector stocks

| Stock | Annual Return (%) | Annual Risk (%) |
|---|---|---|
| ADANIENT | 123.42 | 51.96 |
| TATASTEEL | 18.57 | 38.16 |
| JSWSTEEL | 28.63 | 37.70 |
| HINDALCO | 21.81 | 40.81 |
| VEDL | 13.49 | 45.03 |
| JINDALSTEL | 20.65 | 54.52 |
| APLAPOLLO | 70.37 | 39.81 |
| SAIL | 13.67 | 48.64 |
| HINDZINC | 3.00 | 33.98 |
| NATIONALUM | 18.98 | 44.98 |

Table 37 presents the weights allocation to different stocks for the three portfolio design approaches. HINDZINC receives the highest allocation as per the minimum risk portfolio while ADANIENT receives the highest allocation as per the optimum risk portfolio.

**Table 37.** The portfolio weights for the metal sector stocks

| Stock | EWQ | MRP | MVP/ORP |
|---|---|---|---|
| ADANIENT | 0.1 | 0.089969 | 0.286688 |
| TATASTEEL | 0.1 | 0.125841 | 0.042599 |
| JSWSTEEL | 0.1 | 0.091772 | 0.059989 |
| HINDALCO | 0.1 | 0.020789 | 0.1659 |
| VEDL | 0.1 | 0.047177 | 0.000744 |
| JINDALSTEL | 0.1 | 0.010119 | 0.056485 |
| APLAPOLLO | 0.1 | 0.196066 | 0.251131 |
| SAIL | 0.1 | 0.059842 | 0.015328 |
| HINDZINC | 0.1 | 0.275785 | 0.08656 |
| NATIONALUM | 0.1 | 0.082639 | 0.034577 |

Figure 15 depicts the weight allocation to the metal sector stocks done by the optimum risk portfolio.

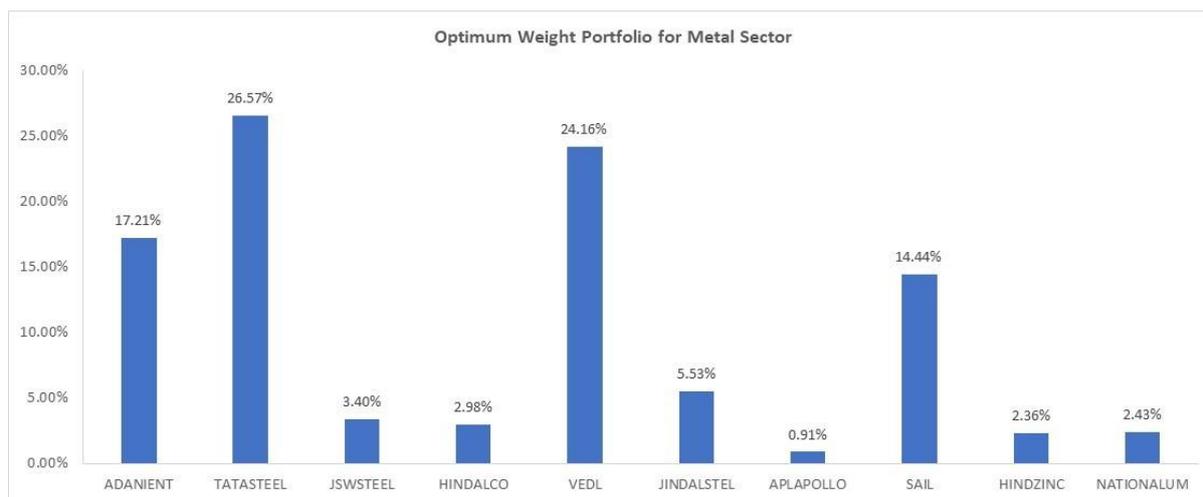

**Figure 15.** The ORP portfolio weights for the metal sector stocks

The risk and return values for the three portfolios, computed using stock prices over the training period, are depicted in Table 38. It is evident that the optimum risk portfolio yields the highest return and the equal weights portfolio exhibits the highest risk.

**Table 38.** The return and risk of the metal sector portfolio

| Metric | EWP | MRP | MVP/ORP |
|---|---|---|---|
| Portfolio annual return(%) | 33.26 | 34.38 | 61.48 |
| Portfolio annual risk (%) | 31.64 | 27.57 | 30.78 |

**Table 39.** The return of the equal-weight portfolio of the metal sector stocks

| | | Date: Jan 3, 2022 | | | Date: Dec 31, 2022 | | RETURN |
|---|---|---|---|---|---|---|---|
| Stock | Weights | Price | Amount Invested | No. of Stock | Price | Value of Stock | |
| ADANIENT | 0.1 | 1717 | 10000 | 5.82 | 3858 | 22469 | |
| TATASTEEL | 0.1 | 114 | 10000 | 87.53 | 113 | 9860 | |
| JSWSTEEL | 0.1 | 667 | 10000 | 14.99 | 768 | 11514 | |
| HINDALCO | 0.1 | 478 | 10000 | 20.92 | 473 | 9902 | 14.38% |
| VEDL | 0.1 | 354 | 10000 | 28.26 | 308 | 8717 | |
| JINDALSTEL | 0.1 | 386 | 10000 | 25.92 | 581 | 15047 | |
| APLAPOLLO | 0.1 | 949 | 10000 | 10.54 | 1092 | 11505 | |
| SAIL | 0.1 | 110 | 10000 | 90.83 | 83 | 7507 | |
| HINDZINC | 0.1 | 320 | 10000 | 31.29 | 322 | 10066 | |
| NATIONALUM | 0.1 | 103 | 10000 | 96.95 | 80 | 7794 | |
| | | | 100000 | | | 114381 | |

The annual return for an investor, investing INR 100,000 on Jan 3, 2022, following the equal weight portfolio approach is shown in Table 39. The investor receives a return of 14.38% at the end of the twelve months.

**Table 40.** The return of the optimum risk portfolio of the metal sector stocks

| Stock | Date: Jan 3, 2022 | | | | Date: Dec 31, 2022 | | RETURN |
|---|---|---|---|---|---|---|---|
| | Weights | Price | Amount Invested | No. of Stock | Price | Value of Stock | |
| ADANIENT | 0.286688 | 1717 | 28668 | 16.7 | 3858 | 64417 | |
| TATASTEEL | 0.042599 | 114 | 4260 | 37.29 | 113 | 4200 | |
| JSWSTEEL | 0.059989 | 667 | 5999 | 8.99 | 768 | 6907 | |
| HINDALCO | 0.1659 | 478 | 16590 | 34.7 | 473 | 16427 | 41.97% |
| VEDL | 0.000744 | 354 | 74 | 0.21 | 308 | 65 | |
| JINDALSTEL | 0.056485 | 386 | 5649 | 14.64 | 581 | 8499 | |
| APLAPOLLO | 0.251131 | 949 | 25113 | 26.46 | 1092 | 28893 | |
| SAIL | 0.015328 | 110 | 1533 | 13.92 | 83 | 1151 | |
| HINDZINC | 0.08656 | 320 | 8656 | 27.08 | 322 | 8713 | |
| NATIONALUM | 0.034577 | 103 | 3458 | 33.52 | 80 | 2695 | |
| | | | 100000 | | | 141967 | |

The performance of the optimum risk portfolio for the metal sector stocks is shown in Table 40. To compare the performance of this portfolio with that of the equal-weight portfolio, the initial amount of investment of INR 100,000 is kept constant. The return yielded by the optimum risk portfolio is found to be 41.97% as depicted in Table 40.

Finally, the efficient frontier for the metal sector portfolios is presented in Figure 16. In Figure 16, the minimum risk portfolio is denoted by the red star, and the optimum risk portfolio is denoted by the green star. The stock prices from Jan 1, 2017, to Dec 31, 2021, are considered in plotting the efficient frontier of the metal sector. It is to be noted that in Figure 16, the x-axis denotes the risk while the y-Axis denotes the return.

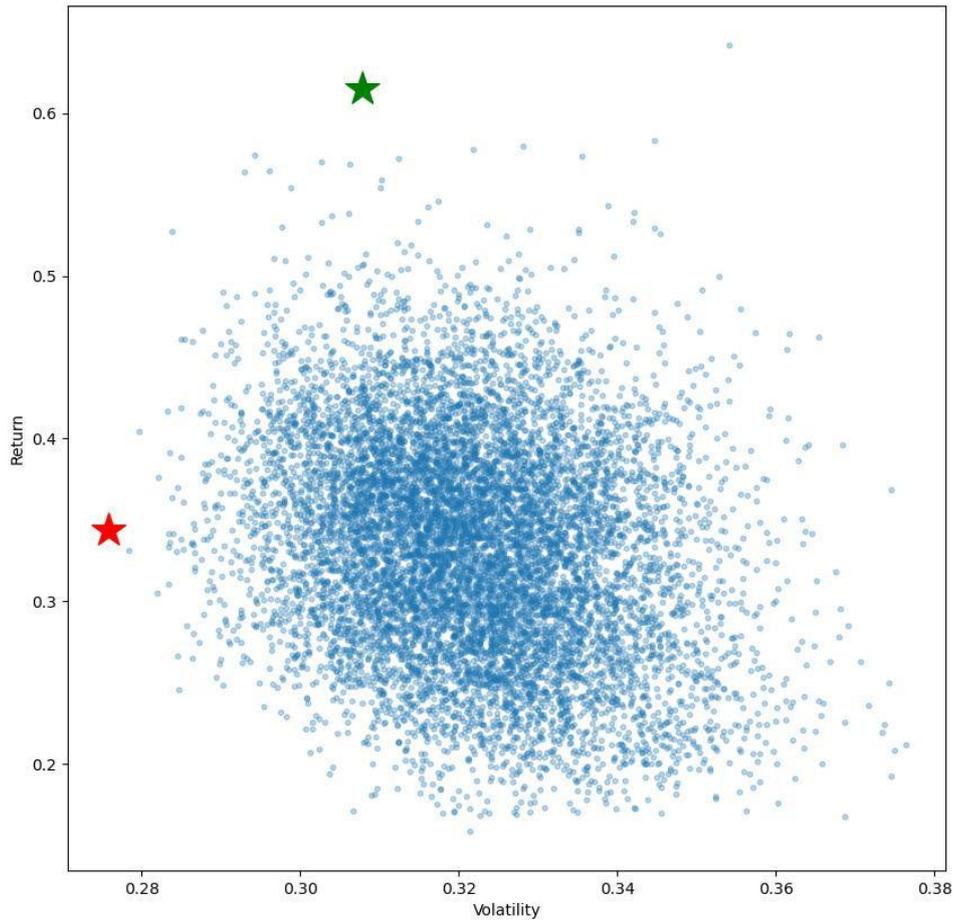

**Figure 12**. The efficient frontier of the metal sector portfolios

### 4.9 The oil & gas sector portfolios

Following are the top ten stocks of this sector based on their free-float market capitalization in the NSE, as per the report released on Dec 30, 2022. The figures below represent the contributions in percentage, of the stocks (computed based on the market capitalization) to the overall index of the oil and gas sector. The ticker name is also mentioned with a pair of parentheses just beside the name of the respective stock.

- i. Reliance Industries (RELIANCE): 31.93
- ii. Adani Total Gas (ATGL): 12.77
- iii. Oil & Natural Gas Corporation (ONGC): 12.44
- iv. Bharat Petroleum Corporation (BPCL): 8.15
- v. Indian Oil Corporation (IOC): 7.58
- vi. GAIL (India) (GAIL): 6.24
- vii. Petronet LNG (PETRONET): 3.96
- viii. Hindustan Petroleum Corporation (HINDPETRO): 3.69
- ix. Indraprastha Gas (IGL): 3.62
- x. Gujarat Gas (GUJGASLTD): 2.03

Table 41 represents the annual return and volatility for oil and gas sector stocks for the training period from Jan 1, 2017, to Dec 31, 2021. ATGL shows the highest annual return while

GAIL yields the lowest. ATGL also exhibits the highest annual risk while PETRONET has the lowest.

Table 41 represents the annual return and volatility for the oil and gas sector stocks for the training period from Jan 1, 2017, to Dec 31, 2021. ADANIENT shows the highest annual return while HINDZINC yields the lowest. JINDALSTEL has the highest annual risk while HINDZINC has the lowest.

**Table 41.** The return and risk of oil and gas sector stocks

| Stock | Annual Return (%) | Annual Risk (%) |
|---|---|---|
| RELIANCE | 27.11 | 31.16 |
| ATGL | 183.56 | 55.93 |
| ONGC | -3.00 | 36.38 |
| BPCL | -3.94 | 38.96 |
| IOC | -10.70 | 32.73 |
| GAIL | -7.50 | 34.34 |
| PETRONET | -3.17 | 28.56 |
| IGL | 12.69 | 31.51 |
| HINDPETRO | -4.63 | 42.24 |
| GUJGASLTD | 46.02 | 33.07 |

Table 42 presents the weights allocation to the different stocks for the three portfolio design approaches. RELIANCE receives the highest allocation as per the minimum risk portfolio while ATGL receives the highest allocation as per the optimum risk portfolio.

**Table 42.** The portfolio weights for the oil and gas sector stocks

| Stock | EWP | MRP | MVP/ORP |
|---|---|---|---|
| RELIANCE | 0.1 | 0.200318 | 0.143897 |
| ATGL | 0.1 | 0.029992 | 0.235186 |
| ONGC | 0.1 | 0.058862 | 0.004412 |
| BPCL | 0.1 | 0.004858 | 0.015196 |
| IOC | 0.1 | 0.116086 | 0.003464 |
| GAIL | 0.1 | 0.055395 | 0.044979 |
| PETRONET | 0.1 | 0.219963 | 0.098012 |
| IGL | 0.1 | 0.085547 | 0.193947 |
| HINDPETRO | 0.1 | 0.039287 | 0.007842 |
| GUJGASLTD | 0.1 | 0.189693 | 0.253066 |

Figure 17 depicts the weight allocation to the oil and gas sector stocks done by the optimum risk portfolio design approach.

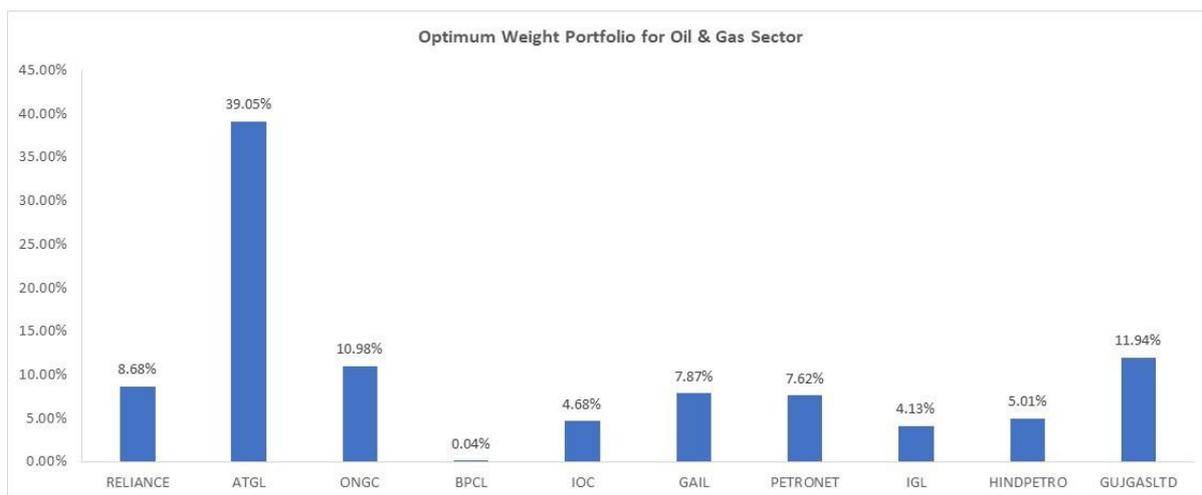

**Figure 17.** The ORP portfolio weights for the oil and gas sector stocks

The risk and return values for the three portfolios, computed using stock prices over the training period, are depicted in Table 43. It is evident that the optimum risk portfolio yields the highest return and it also exhibits the highest risk.

**Table 43.** The return and risk of the oil and gas sector portfolio

| Metric | EWP | MRP | MVP/ORP |
|---|---|---|---|
| Portfolio annual return (%) | 23.64 | 18.02 | 60.38 |
| Portfolio annual risk (%) | 23.11 | 20.39 | 24.23 |

**Table 44.** The return of the equal-weight portfolio of the oil and gas sector stocks

| Stock | Date: Jan 3, 2022 | | | | Date: Dec 31, 2022 | | RETURN |
|---|---|---|---|---|---|---|---|
| | Weights | Price | Amount Invested | No. of Stock | Price | Value of Stock | |
| RELIANCE | 0.1 | 2404 | 10000 | 4.16 | 2547 | 10596 | |
| ATGL | 0.1 | 1744 | 10000 | 5.73 | 3693 | 21177 | |
| ONGC | 0.1 | 143 | 10000 | 69.91 | 147 | 10259 | |
| BPCL | 0.1 | 386 | 10000 | 25.9 | 331 | 8560 | |
| IOC | 0.1 | 75 | 10000 | 132.8 | 77 | 10159 | 14.38% |
| GAIL | 0.1 | 88 | 10000 | 114.11 | 96 | 10960 | |
| PETRONET | 0.1 | 219 | 10000 | 45.7 | 215 | 9847 | |
| IGL | 0.1 | 475 | 10000 | 21.06 | 414 | 8718 | |
| HINDPETRO | 0.1 | 297 | 10000 | 33.68 | 235 | 7920 | |
| GUJGASLTD | 0.1 | 641 | 10000 | 15.6 | 486 | 7584 | |
| | | | 100000 | | | 105780 | |

The annual return for an investor, investing INR 100,000 on Jan 3, 2022, following the equal weight portfolio approach is shown in Table 44. The investor receives a return of 14.38% at the end of the twelve months.

**Table 45.** The return of optimum risk portfolio of the oil and gas sector stocks

| Stock | Date: Jan 3, 2022 | | | | Date: Dec 31, 2022 | | RETURN |
|---|---|---|---|---|---|---|---|
| | Weights | Price | Amount Invested | No. of Stock | Price | Value of Stock | |
| RELIANCE | 0.143897 | 2404 | 14390 | 5.99 | 2547 | 15248 | |
| ATGL | 0.235186 | 1744 | 23519 | 13.49 | 3693 | 49805 | |
| ONGC | 0.004412 | 143 | 441 | 3.08 | 147 | 453 | |
| BPCL | 0.015196 | 386 | 1520 | 3.94 | 331 | 1301 | |
| IOC | 0.003464 | 75 | 346 | 4.6 | 77 | 352 | 18.46% |
| GAIL | 0.044979 | 88 | 4498 | 51.33 | 96 | 4930 | |
| PETRONET | 0.098012 | 219 | 9801 | 44.8 | 215 | 9651 | |
| IGL | 0.193947 | 475 | 19395 | 40.85 | 414 | 16909 | |
| HINDPETRO | 0.007842 | 297 | 784 | 2.64 | 235 | 621 | |
| GUJGASLTD | 0.253066 | 641 | 25306 | 39.48 | 486 | 19193 | |
| | | | 100000 | | | 118463 | |

The performance of the optimum risk portfolio for the oil and gas sector stocks is shown in Table 45. To compare the performance of this portfolio with that of the equal-weight portfolio, the initial amount of investment of INR 100,000 is kept constant. The return yielded by the optimum risk portfolio is found to be 18.46% as depicted in Table 45.

Finally, the efficient frontier for the oil and gas sector portfolios is presented in Figure 18. In Figure 18, the minimum risk portfolio is denoted by the red star, and the optimum risk portfolio is denoted by the green star. The stock prices from Jan 1, 2017, to Dec 31, 2021, are considered in plotting the efficient frontier of the oil and gas sector. It is to be noted that in Figure 18, the x-axis denotes the risk while the y-Axis denotes the return.

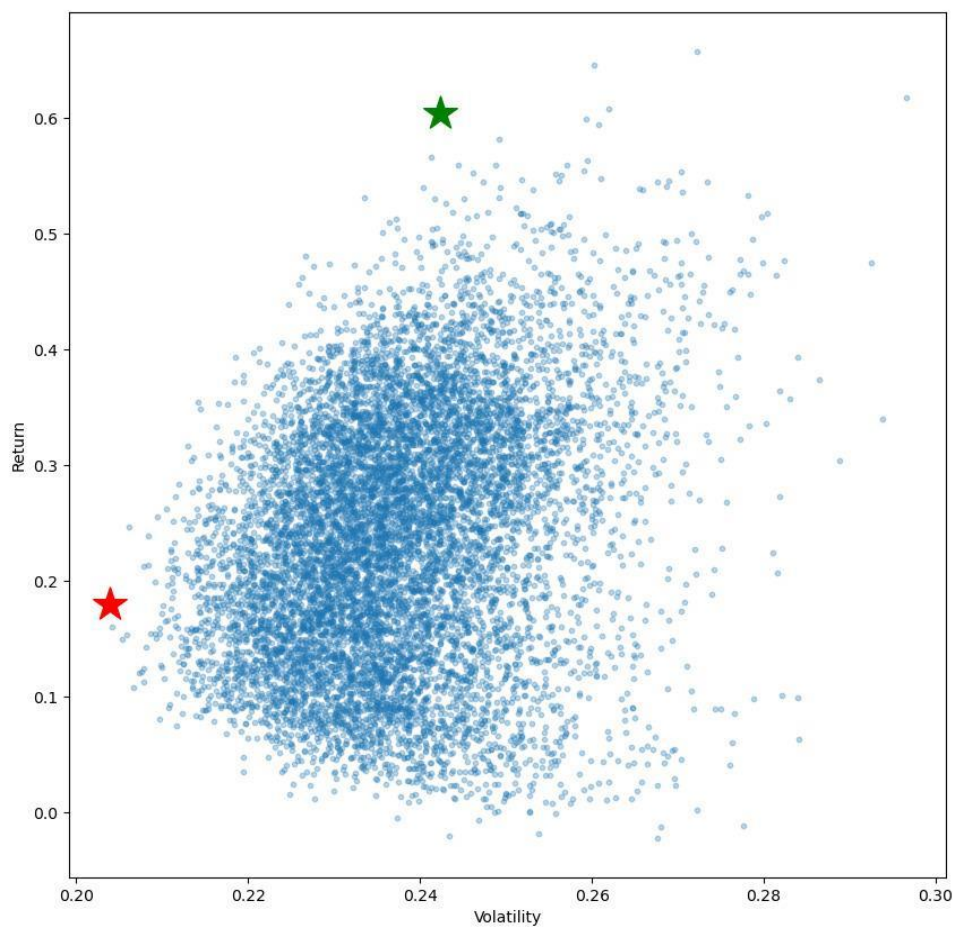

**Figure 18.** The efficient frontier of the oil and gas sector portfolios

### 4.10 The pharma sector portfolios

Following are the top ten stocks of this sector based on their free-float market capitalization in the NSE, as per the report released on Dec 30, 2022. The figures below represent the contributions in percentage, of the stocks (computed based on the market capitalization) to the overall index of the pharma sector. The ticker name is also mentioned with a pair of parentheses just beside the name of the respective stock.

- i. Sun Pharmaceutical Industries (SUNPHARMA): 26.88
- ii. Cipla (CIPLA): 13.25
- iii. Dr. Reddy's Laboratories (DRREDDY): 12.65
- iv. Divi's Laboratories (DIVISLAB): 10.17
- v. Lupin (LUPIN): 4.28
- vi. Alkem Laboratories (ALKEM): 3.55
- vii. Torrent Pharmaceuticals (TORNTPHARM): 3.35
- viii. Laurus Labs (LAURUSLABS): 3.12
- ix. Biocon (BIOCON): 2.80
- x. Aurobindo Pharma (AUROPHARMA): 2.76

Table 46 represents the annual return and volatility for the pharma sector stocks for the training period from Jan 1, 2017, to Dec 31, 2021. LAURUSLABS shows the highest annual return while LUPIN yields the lowest. AUROPHARMA also has the highest annual risk while ALKEM has the lowest.

**Table 46.** The return and risk of the pharma sector stocks

| Stock | Annual Return (%) | Annual Risk (%) |
|---|---|---|
| SUNPHARMA | 13.89 | 32.70 |
| CIPLA | 16.01 | 27.90 |
| DRREDDY | 23.40 | 28.60 |
| DIVISLAB | 47.35 | 33.75 |
| LUPIN | 2.77 | 31.15 |
| LAURUSLABS | 102.03 | 36.24 |
| TORNTPHARM | 24.45 | 29.25 |
| ALKEM | 15.25 | 26.28 |
| AUROPHARMA | 12.56 | 40.51 |
| BIOCON | 11.82 | 34.72 |

Table 47 presents the weights allocation to different stocks for the three portfolio design approaches. ALKEM receives the highest allocation as per the minimum risk portfolio while LAURUSLABS receives the highest allocation as per the optimum risk portfolio.

**Table 47.** The portfolio weights for the pharma sector stocks

| Stock | EWP | MRP | MVP/ORP |
|---|---|---|---|
| SUNPHARMA | 0.1 | 0.088563 | 0.014507 |
| CIPLA | 0.1 | 0.129415 | 0.086986 |
| DRREDDY | 0.1 | 0.104707 | 0.086459 |
| DIVISLAB | 0.1 | 0.063899 | 0.191804 |
| LUPIN | 0.1 | 0.080866 | 0.014662 |
| LAURUSLABS | 0.1 | 0.087915 | 0.281093 |
| TORNTPHARM | 0.1 | 0.118156 | 0.227336 |
| ALKEM | 0.1 | 0.211195 | 0.080163 |
| AUROPHARMA | 0.1 | 0.000819 | 0.004413 |
| BIOCON | 0.1 | 0.114465 | 0.012578 |

Figure 19 depicts the weight allocation to the pharma sector stocks done by the optimum risk portfolio.

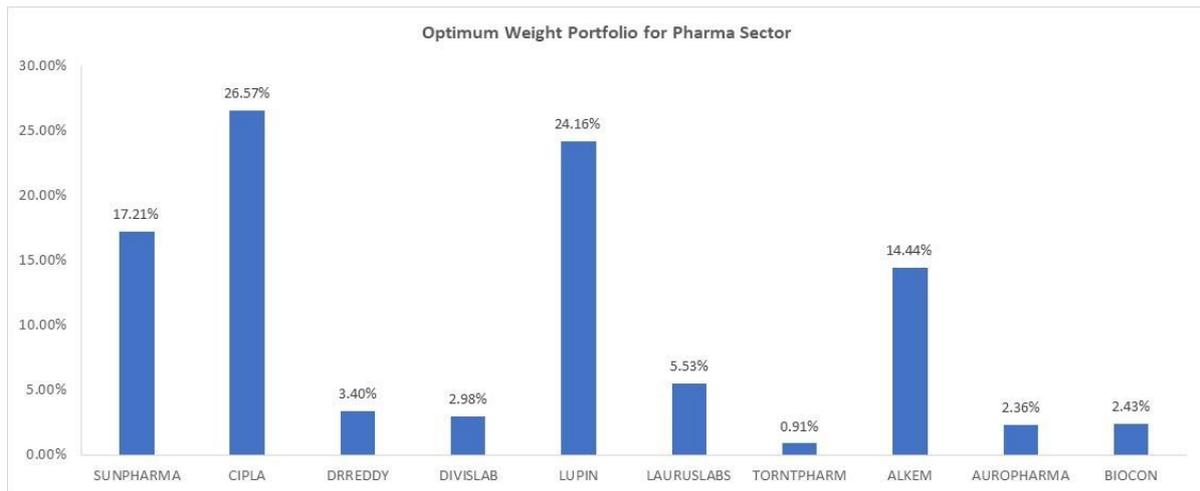

**Figure 13.** The ORP portfolio weights for the pharma sector stocks

The risk and return values for the three portfolios, computed using stock prices over the training period, are depicted in Table 48. It is evident that the optimum risk portfolio yields the highest return and it also exhibits the highest risk.

**Table 48.** The return and risk of the pharma sector portfolios

| Metric | EWP | MRP | MVP/ORP |
|---|---|---|---|
| Portfolio annual return (%) | 26.95 | 25.44 | 48.40 |
| Portfolio annual risk (%) | 20.17 | 18.53 | 20.80 |

**Table 49.** The return of the equal-weight portfolio of the pharma sector stocks

| | | Date: Jan 3, 2022 | | | Date: Dec 31, 2022 | | RETURN |
|---|---|---|---|---|---|---|---|
| Stock | Weights | Price | Amount Invested | No. of Stock | Price | Value of Stock | |
| SUNPHARMA | 0.1 | 849 | 10000 | 11.78 | 1001 | 11796 | |
| CIPLA | 0.1 | 931 | 10000 | 10.75 | 1076 | 11563 | |
| DRREDDY | 0.1 | 4853 | 10000 | 2.06 | 4238 | 8732 | |
| DIVISLAB | 0.1 | 4651 | 10000 | 2.15 | 3413 | 7338 | -14.72% |
| LUPIN | 0.1 | 945 | 10000 | 10.58 | 734 | 7762 | |
| LAURUSLABS | 0.1 | 529 | 10000 | 18.91 | 375 | 7097 | |
| TORNTPHARM | 0.1 | 1626 | 10000 | 6.15 | 1551 | 9536 | |
| ALKEM | 0.1 | 3641 | 10000 | 2.75 | 3005 | 8252 | |
| AUROPHARMA | 0.1 | 732 | 10000 | 13.66 | 438 | 5987 | |
| BIOCON | 0.1 | 363 | 10000 | 27.57 | 262 | 7218 | |
| | | | 100000 | | | 85281 | |

The annual return for an investor, investing INR 100,000 on Jan 3, 2022, following the equal weight portfolio approach is shown in Table 49. The annual return in this case is -14.72%. The negative value indicates that the investor has incurred a loss.

**Table 50.** The return of the optimum risk portfolio of the pharma sector stocks

| Stock | Date: Jan 3, 2022 | | | | Date: Dec 31, 2022 | | RETURN |
|---|---|---|---|---|---|---|---|
| | Weights | Price | Amount Invested | No. of Stock | Price | Value of Stock | |
| SUNPHARMA | 0.014507 | 849 | 1451 | 1.71 | 1001 | 1711 | |
| CIPLA | 0.086986 | 931 | 8699 | 9.35 | 1076 | 10058 | |
| DRREDDY | 0.086459 | 4853 | 8646 | 1.78 | 4238 | 7549 | |
| DIVISLAB | 0.191804 | 4651 | 19180 | 4.12 | 3413 | 14075 | -16.05% |
| LUPIN | 0.014662 | 945 | 1466 | 1.55 | 734 | 1138 | |
| LAURUSLABS | 0.281093 | 529 | 28109 | 53.16 | 375 | 19949 | |
| TORNTPHARM | 0.227336 | 1626 | 22734 | 13.98 | 1551 | 21679 | |
| ALKEM | 0.080163 | 3641 | 8016 | 2.2 | 3005 | 6615 | |
| AUROPHARMA | 0.004413 | 732 | 441 | 0.6 | 438 | 264 | |
| BIOCON | 0.012578 | 363 | 1258 | 3.47 | 262 | 908 | |
| | | | 100000 | | | 83946 | |

The performance of the optimum risk portfolio for the pharma sector stocks is shown in Table 50. To compare the performance of this portfolio with that of the equal-weight portfolio, the initial amount of investment of INR 100,000 is kept constant. The return of the optimum risk portfolio over the same period is found to be -16.05%. The investor has incurred a loss using the optimum risk portfolio for the pharma sector.

Finally, the efficient frontier for the pharma sector portfolios is presented in Figure 20. In Figure 20, the minimum risk portfolio is denoted by the red star, and the optimum risk portfolio is denoted by the green star. The stock prices from Jan 1, 2017, to Dec 31, 2021, are considered in plotting the efficient frontier of the pharma sector. It is to be noted that in Figure 20, the x-axis denotes the risk while the y-axis denotes the return.

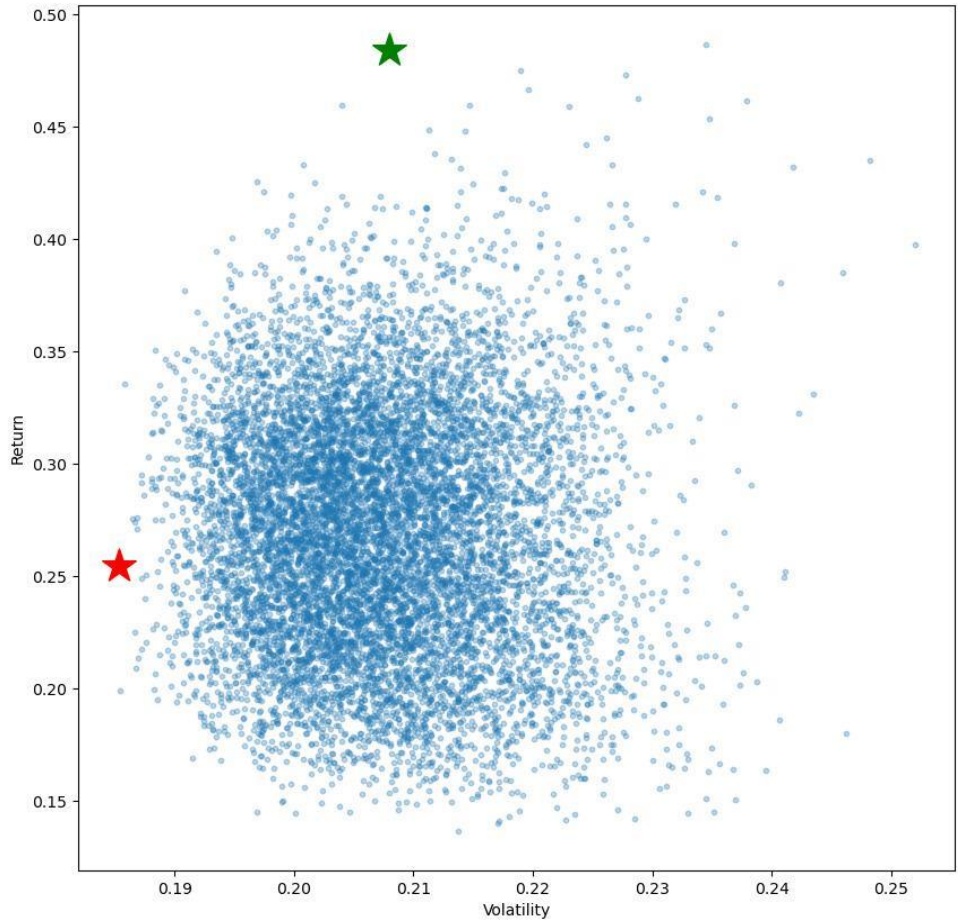

**Figure 20**. The efficient frontier of the pharma sector portfolios

### 4.11 The Public Sector Unit (PSU) banks portfolios

Following are the top ten stocks of this sector based on their free-float market capitalization in the NSE, as per the report released on Dec 30, 2022. The figures below represent the contributions in percentage, of the stocks (computed based on the market capitalization) to the overall index of the PSU Bank sector. The ticker name is also mentioned with a pair of parentheses just beside the name of the respective stock.

- i. State Bank of India (SBIN): 29.00
- ii. Bank of Baroda (BANKBARODA): 19.56
- iii. Canara Bank (CANBK): 12.77
- iv. Punjab National Bank (PNB): 12.75
- v. Union Bank of India (UNIONBANK): 7.59
- vi. Indian Bank (INDIANB): 6.31
- vii. Bank of India (BANKINDIA): 5.36
- viii. Indian Overseas Bank (IOB): 1.79
- ix. Bank of Maharashtra (MAHABANK): 1.53
- x. Central Bank of India (CENTRALBK): 1.52

Table 51 represents the annual return and volatility for the training period from Jan 1, 2017, to Dec 31, 2021. SBIN shows the highest annual return while PNB yields the lowest. CANBK has the highest annual risk while SBIN has the lowest.

Table 51 represents the annual return and volatility for the PSU bank sector stocks for the training period from Jan 1, 2017, to Dec 31, 2021. SBIN shows the highest annual return while PNB yields the lowest. CENTRALBK has the highest annual risk while SBIN has the lowest.

**Table 51.** The return and risk of the PSU bank sector stocks

| Stock | Annual Return (%) | Annual Risk (%) |
|---|---|---|
| SBIN | 14.53 | 36.87 |
| BANKBARODA | -11.65 | 45.39 |
| CANBK | -7.47 | 46.69 |
| PNB | -26.96 | 46.35 |
| UNIONBANK | -20.39 | 46.36 |
| INDIANB | -11.53 | 49.46 |
| BANKINDIA | -24.08 | 47.39 |
| IOB | 7.03 | 44.29 |
| MAHABANK | 0.02 | 46.49 |
| CENTRALBK | -17.39 | 49.63 |

Table 52 presents weights allocation to different stocks for the three portfolio design approaches. SBIN receives the highest allocation as per the minimum risk portfolio as well as the optimum risk portfolio.

**Table 52.** The portfolio weights for the PSU bank sector stocks

| Stock | EWP | MRP | MVP/ORP |
|---|---|---|---|
| SBIN | 0.1 | 0.213164 | 0.254636 |
| BANKBARODA | 0.1 | 0.005341 | 0.040239 |
| CANBK | 0.1 | 0.033512 | 0.23961 |
| PNB | 0.1 | 0.124246 | 0.05054 |
| UNIONBANK | 0.1 | 0.034152 | 0.000181 |
| INDIANB | 0.1 | 0.073182 | 0.061639 |
| BANKINDIA | 0.1 | 0.026572 | 0.013106 |
| IOB | 0.1 | 0.120369 | 0.249011 |
| MAHABANK | 0.1 | 0.194339 | 0.090445 |
| CENTRALBK | 0.1 | 0.175122 | 0.000593 |

Figure 21 depicts the weight allocation for the PSU bank sector stocks done by the optimum risk portfolio.

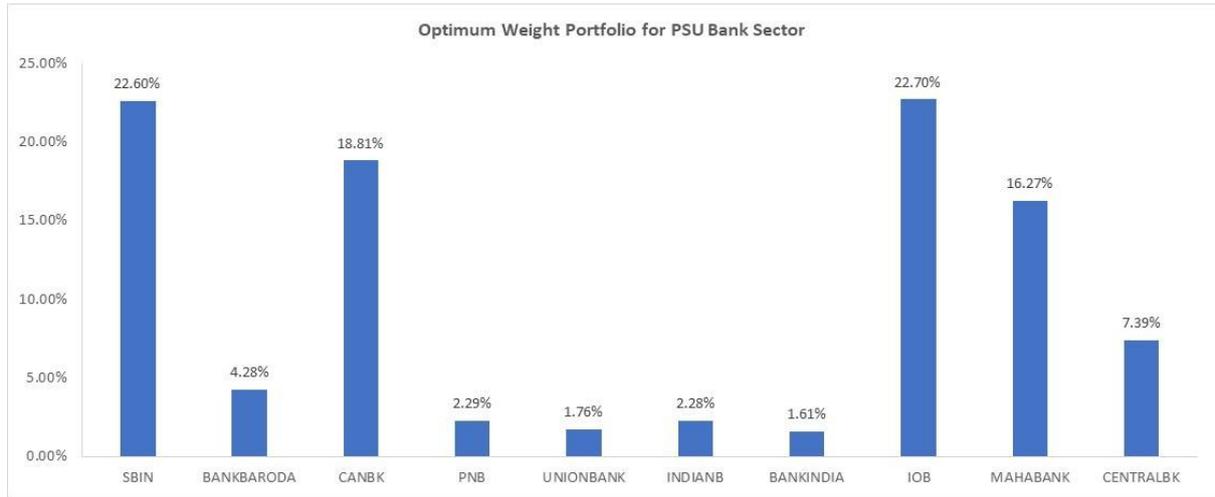

**Figure 14.** The ORP portfolio weights for the PSU banking sector stocks

The risk and return values for the three portfolios, computed using stock prices over the training period, are depicted in Table 53. It is evident that the optimum risk portfolio yields the highest return and the equal weights portfolio exhibits the highest risk.

**Table 53.** The return and risk of the PSU bank sector portfolios

| Metric | EWP | MRP | MVP/ORP |
|---|---|---|---|
| Portfolio annual return (%) | -9.79 | -4.94 | 0.79 |
| Portfolio annual risk (%) | 35.98 | 34.12 | 34.78 |

**Table 54.** The return of the equal-weight portfolio of the PSU bank sector stocks

| Stock | Date: Jan 3, 2022 | | | | Date: Dec 31, 2022 | | RETURN |
|---|---|---|---|---|---|---|---|
| | Weights | Price | Amount Invested | No. of Stock | Price | Value of Stock | |
| SBIN | 0.1 | 471 | 10000 | 21.24 | 614 | 13035 | |
| BANKBARODA | 0.1 | 84 | 10000 | 119.33 | 186 | 22160 | |
| CANBK | 0.1 | 205 | 10000 | 48.67 | 333 | 16223 | |
| PNB | 0.1 | 38 | 10000 | 263.16 | 56 | 14855 | 67.70% |
| UNIONBANK | 0.1 | 44 | 10000 | 226.5 | 80 | 18211 | |
| INDIANB | 0.1 | 142 | 10000 | 70.35 | 285 | 20074 | |
| BANKINDIA | 0.1 | 53 | 10000 | 189.93 | 88 | 16762 | |
| IOB | 0.1 | 21 | 10000 | 486.62 | 32 | 15620 | |
| MAHABANK | 0.1 | 19 | 10000 | 515.46 | 31 | 15773 | |
| CENTRALBK | 0.1 | 21 | 10000 | 466.2 | 32 | 14988 | |
| | | | 100000 | | | 167701 | |

The annual return for an investor, investing INR 100,000 on Jan 3, 2022, and following the equal weight portfolio approach is shown in Table 54. The investor receives a return of 67.70% at the end of the twelve months.

**Table 55.** The return of the optimum risk portfolio of the PSU bank sector stocks

| Stock | Date: Jan 3, 2022 | | | | Date: Dec 31, 2022 | | RETURN |
|---|---|---|---|---|---|---|---|
| | Weights | Price | Amount Invested | No. of Stock | Price | Value of Stock | |
| SBIN | 0.254636 | 471 | 25464 | 54.09 | 614 | 33192 | |
| BANKBARODA | 0.040239 | 84 | 4024 | 48.02 | 186 | 8917 | |
| CANBK | 0.23961 | 205 | 23961 | 116.63 | 333 | 38872 | |
| PNB | 0.05054 | 38 | 5054 | 133 | 56 | 7508 | |
| UNIONBANK | 0.000181 | 44 | 18 | 0.41 | 80 | 33 | 56.34% |
| INDIANB | 0.061639 | 142 | 6164 | 43.36 | 285 | 12373 | |
| BANKINDIA | 0.013106 | 53 | 1311 | 24.89 | 88 | 2197 | |
| IOB | 0.249011 | 21 | 24900 | 1211.73 | 32 | 38897 | |
| MAHABANK | 0.090445 | 19 | 9045 | 466.21 | 31 | 14266 | |
| CENTRALBK | 0.000593 | 21 | 59 | 2.76 | 32 | 89 | |
| | | | 100000 | | | 156344 | |

The performance of the optimum risk portfolio for the PSU bank sector stocks is shown in Table 55. To compare the performance of this portfolio with that of the equal-weight portfolio, the initial amount of investment of INR 100,000 is kept constant. The return yielded by the optimum risk portfolio is found to be 56.34% as depicted in Table 55.

Finally, the efficient frontier for the PSU bank sector portfolios is presented in Figure 22. In Figure 22, the minimum risk portfolio is denoted by the red star, and the optimum risk portfolio is denoted by the green star. The stock prices from Jan 1, 2017, to Dec 31, 2021, are considered in plotting the efficient frontier of the PSU banks sector. It is to be noted that in Figure 22, the x-axis denotes the risk while the y-Axis denotes the return.

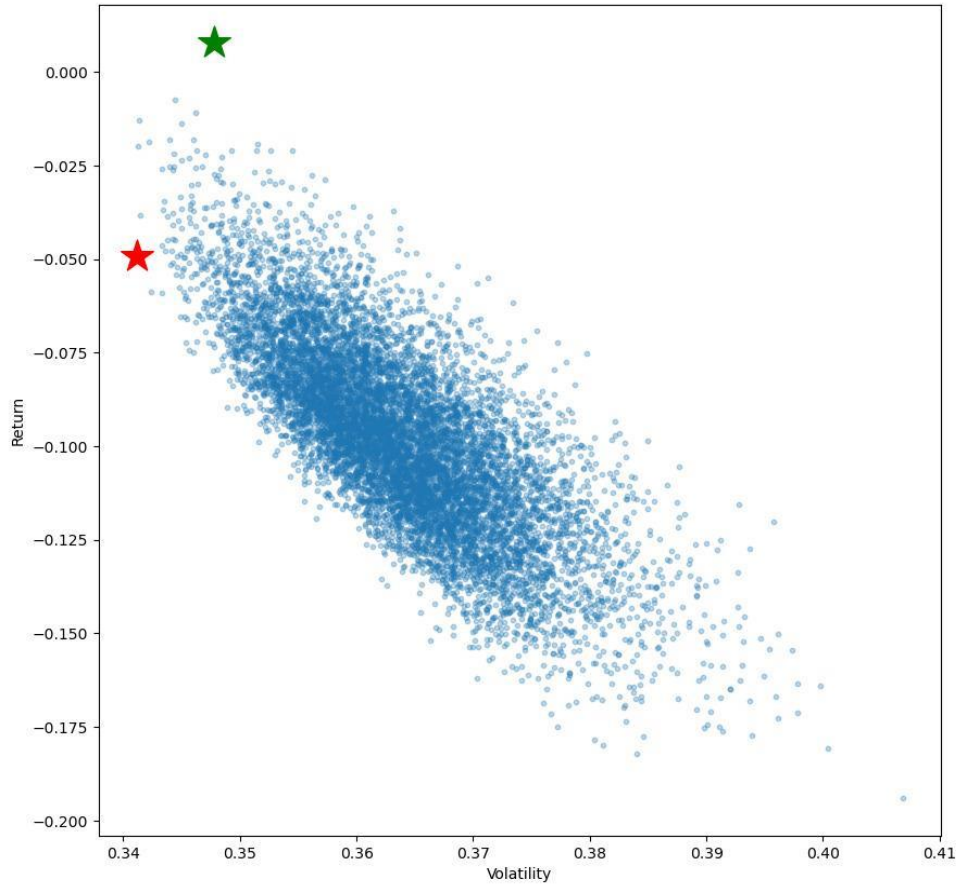

**Figure 15.** The efficient frontier of the PSU bank sector portfolio

### 4.12 The private (PVT) banks sector portfolios

Following are the top ten stocks of this sector based on their free-float market capitalization in the NSE, as per the report released on Dec 30, 2022. The figures below represent the contributions in percentage, of the stocks (computed based on the market capitalization) to the overall index of the PVT bank sector. The ticker name is also mentioned with a pair of parentheses just beside the name of the respective stock.

  i.   HDFC Bank (HDFCBANK): 26.83
 ii.   ICICI Bank (ICICIBANK): 24.58
iii.   Kotak Mahindra Bank (KOTAKBANK): 10.63
 iv.   Axis Bank (AXISBANK): 10.45
  v.   IndusInd Bank (INDUSINDBK): 10.07
 vi.   Federal Bank (FEDERALBNK): 5.81
vii.   Bandhan Bank (BANDHANBNK): 3.78
viii.  IDFC First Bank (IDFCFIRSTB): 3.63
 ix.   City Union Bank (CUB): 2.33
  x.   RBL Bank (RBLBANK): 1.89

Table 56 represents the annual return and volatility for the PVT bank sector stocks for the training period from Jan 1, 2017, to Dec 31, 2021. ICICIBANK produces the highest annual return while RBLBANK yields the lowest return. BANDHANBNK exhibits the highest annual risk while HDFCBANK exhibits the lowest volatility.

**Table 56.** The return and risk of PVT bank sector stocks

| Stock | Annual Return (%) | Annual Risk (%) |
|---|---|---|
| HDFCBANK | 12.28 | 24.91 |
| ICICIBANK | 25.49 | 35.63 |
| AXISBANK | 5.80 | 38.54 |
| KOTAKBANK | 16.73 | 29.28 |
| INDUSINDBK | -12.52 | 48.43 |
| FEDERALBNK | -4.87 | 40.64 |
| IDFCFIRSTB | -0.85 | 41.21 |
| BANDHANBNK | -21.88 | 56.42 |
| CUB | -2.25 | 34.05 |
| RBLBANK | -26.30 | 49.44 |

Table 57 presents the weights allocation to different stocks for the three portfolio design approaches. HDFCBANK receives the highest allocation as per the minimum risk portfolio while ICICIBANK receives the highest allocation as per the optimum risk portfolio.

**Table 57.** The portfolio weights for the PVT bank sector

| Stock | EWP | MRP | MVP/ORP |
|---|---|---|---|
| HDFCBANK | 0.1 | 0.325157 | 0.204524 |
| ICICIBANK | 0.1 | 0.045523 | 0.264651 |
| AXISBANK | 0.1 | 0.061599 | 0.070239 |
| KOTAKBANK | 0.1 | 0.155792 | 0.230697 |
| INDUSINDBK | 0.1 | 0.0113 | 0.035574 |
| FEDERALBNK | 0.1 | 0.000128 | 0.028255 |
| IDFCFIRSTB | 0.1 | 0.096205 | 0.026738 |
| BANDHANBNK | 0.1 | 0.056141 | 0.045185 |
| CUB | 0.1 | 0.195319 | 0.059839 |
| RBLBANK | 0.1 | 0.052836 | 0.034298 |

Figure 23 depicts the weight allocation to the PVT bank sector stocks done by the optimum risk portfolio.

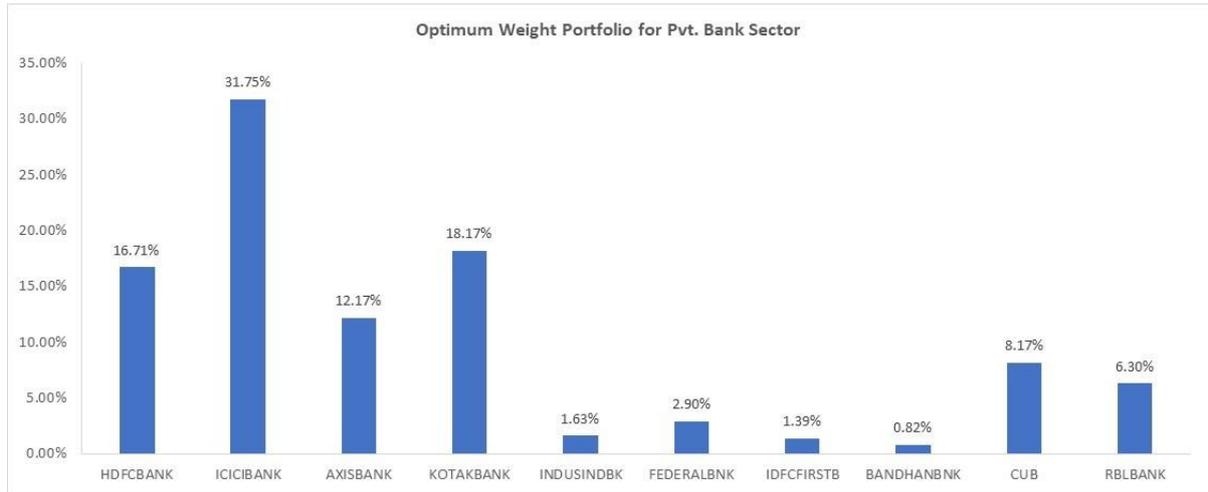

**Figure 16.** The ORP portfolio weights for the PVT bank sector stocks

The risk and return values for the three portfolios, computed using the stock prices over the training period, are depicted in Table 58. It is evident that the optimum risk portfolio yields the highest return and the equal weights portfolio exhibits the highest risk.

**Table 58.** The return and risk of PVT bank sector portfolio

| Metric | EWP | MRP | MVP/ORP |
|---|---|---|---|
| Portfolio annual return (%) | -0.84 | 4.84 | 10.89 |
| Portfolio annual risk (%) | 29.05 | 24.22 | 25.75 |

**Table 59.** The return of the equal-weight portfolio of the PVT bank sector stocks

| Stock | Date: Jan 3, 2022 | | | | Date: Dec 31, 2022 | | RETURN |
|---|---|---|---|---|---|---|---|
| | Weights | Price | Amount Invested | No. of Stock | Price | Value of Stock | |
| HDFCBANK | 0.1 | 1520 | 10000 | 6.58 | 1628 | 10714 | |
| ICICIBANK | 0.1 | 765 | 10000 | 13.08 | 891 | 11650 | |
| AXISBANK | 0.1 | 696 | 10000 | 14.36 | 934 | 13409 | |
| KOTAKBANK | 0.1 | 1824 | 10000 | 5.48 | 1827 | 10015 | 22.76% |
| INDUSINDBK | 0.1 | 912 | 10000 | 10.96 | 1220 | 13374 | |
| FEDERALBNK | 0.1 | 87 | 10000 | 114.68 | 139 | 15946 | |
| IDFCFIRSTB | 0.1 | 50 | 10000 | 201.41 | 59 | 11843 | |
| BANDHANBNK | 0.1 | 252 | 10000 | 39.62 | 234 | 9279 | |
| CUB | 0.1 | 139 | 10000 | 72.15 | 180 | 13016 | |
| RBLBANK | 0.1 | 133 | 10000 | 75.33 | 179 | 13514 | |
| | | | 100000 | | | 122760 | |

The annual return for an investor, investing INR 100,000 on Jan 3, 2022, and following the equal-weight portfolio approach is shown in Table 59. The investor receives a return of 22.76% at the end of the twelve months.

**Table 60.** The return of the optimum risk portfolio of the PVT bank sector stocks

| Stock | Date: Jan 3, 2022 | | | | Date: Dec 31, 2022 | | RETURN |
|---|---|---|---|---|---|---|---|
| | Weights | Price | Amount Invested | No. of Stock | Price | Value of Stock | |
| HDFCBANK | 0.204524 | 1520 | 20452 | 13.46 | 1628 | 21913 | |
| ICICIBANK | 0.264651 | 765 | 26465 | 34.61 | 891 | 30831 | |
| AXISBANK | 0.070239 | 696 | 7024 | 10.09 | 934 | 9418 | |
| KOTAKBANK | 0.230697 | 1824 | 23070 | 12.64 | 1827 | 23105 | 14.31% |
| INDUSINDBK | 0.035574 | 912 | 3557 | 3.9 | 1220 | 4758 | |
| FEDERALBNK | 0.028255 | 87 | 2826 | 32.4 | 139 | 4506 | |
| IDFCFIRSTB | 0.026738 | 50 | 2674 | 53.85 | 59 | 3167 | |
| BANDHANBNK | 0.045185 | 252 | 4518 | 17.9 | 234 | 4193 | |
| CUB | 0.059839 | 139 | 5984 | 43.17 | 180 | 7789 | |
| RBLBANK | 0.034298 | 133 | 3430 | 25.84 | 179 | 4635 | |
| | | | 100000 | | | 114315 | |

The performance of the optimum risk portfolio for the PVT bank sector stocks is shown in Table 60. To compare the performance of this portfolio with that of the equal-weight portfolio, the initial amount of investment of INR 100,000 is kept constant. The return yielded by the optimum risk portfolio is found to be 14.31% as depicted in Table 60.

Finally, the efficient frontier for the PVT bank sector portfolios is presented in Figure 24. In Figure 24, the minimum risk portfolio is denoted by the red star, and the optimum risk portfolio is denoted by the green star. The stock prices from Jan 1, 2017, to Dec 31, 2021, are considered in plotting the efficient frontier of the PVT banks sector. It is to be noted that in Figure 24, the x-axis denotes the risk while the y-Axis denotes the return.

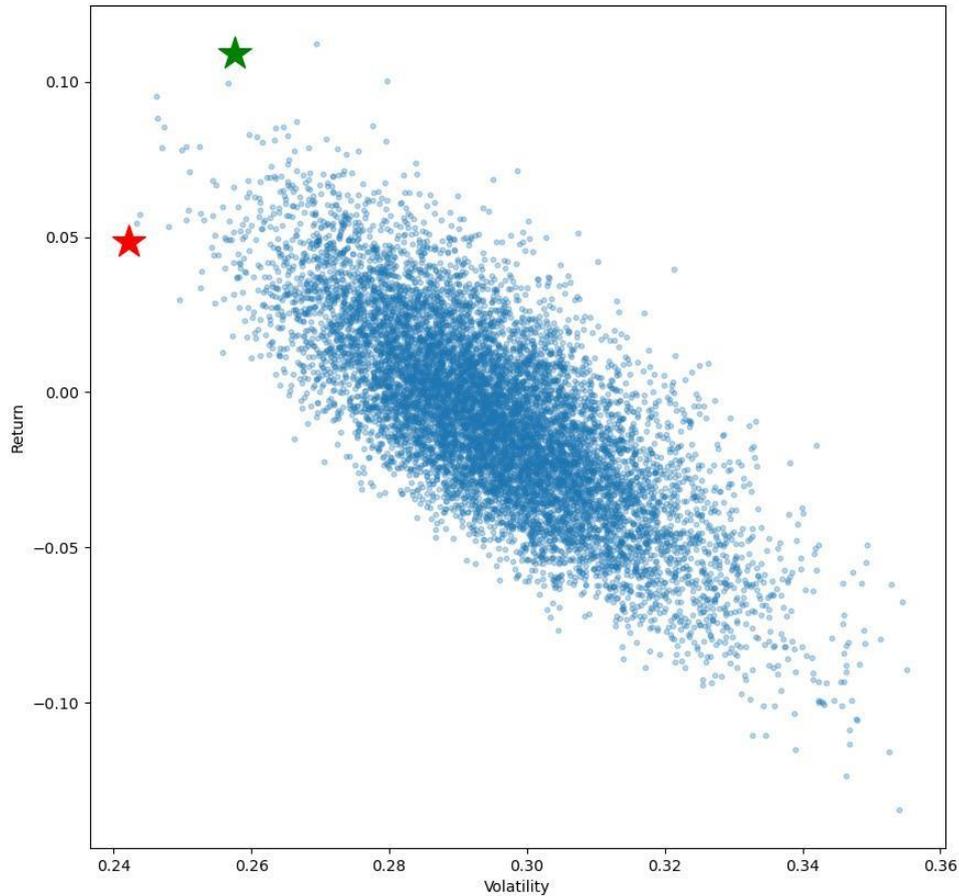

**Figure 17.** The efficient frontier of the PVT bank sector portfolio

### 4.13 The realty sector portfolio

Following are the top ten stocks of this sector based on their free-float market capitalization in the NSE, as per the report released on Dec 30, 2022. The figures below represent the contributions in percentage, of the stocks (computed based on the market capitalization) to the overall index of the realty sector. The ticker name is also mentioned with a pair of parentheses just beside the name of the respective stock.

    i.       DLF (DLF): 25.45
    ii.      Godrej Properties (GODREJPROP): 15.31
    iii.     Phoenix Mills (PHOENIXLTD): 14.67
    iv.     Oberoi Realty (OBEROIRLTY): 11.67
    v.      Macrotech Developers (LODHA): 11.55
    vi.     Prestige Estate Projects (PRESTIGE): 7.37
    vii.    Brigade Enterprises (BRIGADE): 6.66
    viii.   Indiabulls Real Estate (IBREALEST): 3.21
    ix.     Sobha (SOBHA): 2.43
    x.      Sunteck Realty (SUNTECK): 1.70

The stock of LODHA was found to contain more than 30% of its observations missing during the training period from Jan 1, 2017, to Dec 31, 2021. The first observation of this stock was available on April 19, 2021. Hence, for LODHA, only 15% of the observations were available. The computation for the portfolios for the realty sector, therefore, did not involve the stock of LODHA. The remaining nine stocks are used in the portfolio design

Table 61 represents the annual return and volatility for the realty sector stocks for the training period from Jan 1, 2017, to Dec 31, 2021. BRIGADE shows the highest annual return while SUNTECK yields the lowest. IBREALEST has the highest annual risk while PHOENIXLTD has the lowest.

**Table 61.** The return and risk of the realty sector stocks

| Stock | Annual Return (%) | Annual Risk (%) |
|---|---|---|
| DLF | 16.76% | 45.67% |
| GODREJPROP | 30.32% | 43.43% |
| PHOENIXLTD | 14.47% | 39.63% |
| OBEROIRLTY | 17.41% | 41.60% |
| PRESTIGE | 20.01% | 49.23% |
| BRIGADE | 32.60% | 41.70% |
| IBREALEST | 7.88% | 61.60% |
| SOBHA | 20.24% | 46.99% |
| SUNTECK | 6.98% | 42.84% |

Table 62 presents the weights allocation to different stocks for the three portfolio design approaches. PHOENIXLTD receives the highest allocation as per the minimum risk portfolio while SOBHA receives the highest allocation as per the optimum risk portfolio.

**Table 62.** The portfolio weights for the realty sector stocks

| Stock | EWP | MRP | MVP/ORP |
|---|---|---|---|
| DLF | 0.1 | 0.036914 | 0.000519 |
| GODREJPROP | 0.1 | 0.127042 | 0.227038 |
| PHOENIXLTD | 0.1 | 0.249625 | 0.174964 |
| OBEROIRLTY | 0.1 | 0.125977 | 0.080031 |
| PRESTIGE | 0.1 | 0.056054 | 0.007363 |
| BRIGADE | 0.1 | 0.153627 | 0.233699 |
| IBREALEST | 0.1 | 0.034577 | 0.002699 |
| SOBHA | 0.1 | 0.029097 | 0.263355 |
| SUNTECK | 0.1 | 0.187087 | 0.010333 |

Figure 25 depicts the weight allocation to the realty sector stocks done by the optimum risk portfolio.

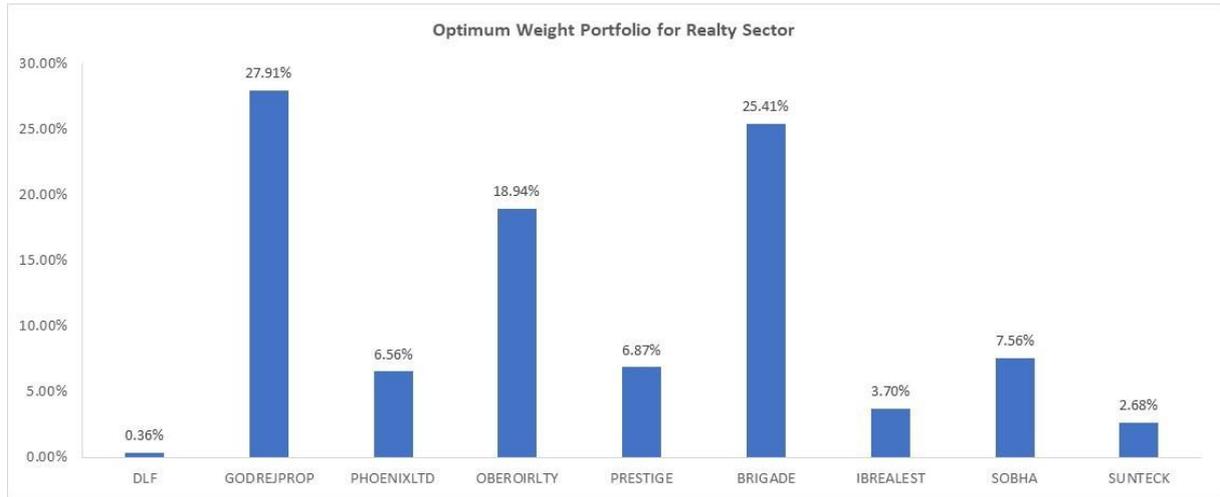

**Figure 18.** The ORP portfolio weights for the realty sector stocks

The risk and return values for the three portfolios, computed using stock prices over the training period, are depicted in Table 63. It is evident that the optimum risk portfolio yields the highest return and it also exhibits the highest risk.

**Table 63.** The return and risk of the realty sector portfolios

| Metric | EWP | MRP | MVP/ORP |
|---|---|---|---|
| Portfolio annual return (%) | 16.67 | 18.57 | 24.01 |
| Portfolio annual risk (%) | 25.70 | 26.82 | 28.67 |

The annual return for an investor, investing INR 100,000 on 3rd January 2022 following the equal weight portfolio approach is shown in Table 64. The annual return in this case is -13.83%. The negative value indicates that the investor has incurred a loss.

**Table 64.** The return of the equal-weight portfolio of the realty sector stocks

| Stock | Date: Jan 3, 2022 | | | | Date: Dec 31, 2022 | | RETURN |
|---|---|---|---|---|---|---|---|
| | Weights | Price | Amount Invested | No. of Stock | Price | Value of Stock | |
| DLF | 0.1 | 395 | 10000 | 25.33 | 375 | 9500 | |
| GODREJPROP | 0.1 | 1904 | 10000 | 5.25 | 1225 | 6434 | |
| PHOENIXLTD | 0.1 | 976 | 10000 | 10.24 | 1423 | 14571 | |
| OBEROIRLTY | 0.1 | 888 | 10000 | 11.26 | 868 | 9771 | -13.83% |
| PRESTIGE | 0.1 | 472 | 10000 | 21.17 | 464 | 9812 | |
| BRIGADE | 0.1 | 494 | 10000 | 20.24 | 465 | 9413 | |
| IBREALEST | 0.1 | 163 | 10000 | 61.54 | 81 | 4991 | |
| SOBHA | 0.1 | 887 | 10000 | 11.27 | 576 | 6492 | |
| SUNTECK | 0.1 | 502 | 10000 | 19.93 | 330 | 6572 | |
| | | | 90000 | | | 77556 | |

The performance of the optimum risk portfolio for the realty sector stocks is shown in Table 65. To compare the performance of this portfolio with that of the equal-weight portfolio, the initial amount of investment of INR 100,000 is kept constant. The return yielded by the optimum risk portfolio is found to be -11.40%, indicating a loss for the investor.

**Table 65.** The return of the optimum risk portfolio of the realty sector stocks

| Stock | Date: Jan 3, 2022 | | | | Date: Dec 31, 2022 | | RETURN |
| --- | --- | --- | --- | --- | --- | --- | --- |
| | Weights | Price | Amount Invested | No. of Stock | Price | Value of Stock | |
| DLF | 0.000519 | 395 | 52 | 0.13 | 375 | 49 | |
| GODREJPROP | 0.227038 | 1904 | 22704 | 11.93 | 1225 | 14607 | |
| PHOENIXLTD | 0.174964 | 976 | 17496 | 17.92 | 1423 | 25494 | |
| OBEROIRLTY | 0.080031 | 888 | 8003 | 9.01 | 868 | 7820 | -11.40% |
| PRESTIGE | 0.007363 | 472 | 736 | 1.56 | 464 | 722 | |
| BRIGADE | 0.233699 | 494 | 23370 | 47.3 | 465 | 21998 | |
| IBREALEST | 0.002699 | 163 | 270 | 1.66 | 81 | 135 | |
| SOBHA | 0.263355 | 887 | 26336 | 29.68 | 576 | 17098 | |
| SUNTECK | 0.010333 | 502 | 1033 | 2.06 | 330 | 679 | |
| | | | 100000 | | | 88602 | |

Finally, the efficient frontier for the realty sector portfolios is presented in Figure 26. In Figure 26, the minimum risk portfolio is denoted by the red star, and the optimum risk portfolio is denoted by the green star. The stock prices from Jan 1, 2017, to Dec 31, 2021, are considered in plotting the efficient frontier of the realty sector. It is to be noted that in Figure 26, the x-axis denotes the risk while the y-axis denotes the return.

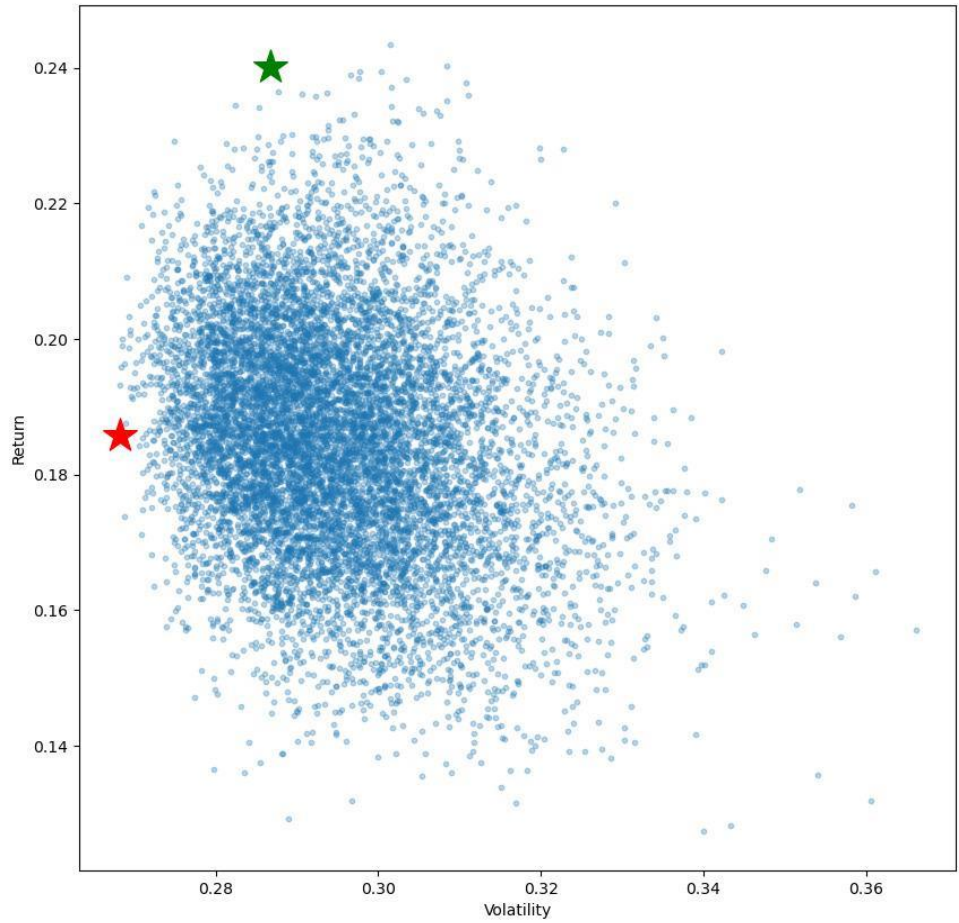

**Figure 26**. The efficient frontier of the realty sector portfolios

## 4.14 Summary of the results

Table 66 presents a summary of the performance results of the portfolios based on their annual returns over the test period. It is observed that the equal-weight portfolio for the PSU banks yielded the highest return, among all. On the other hand, for the IT sector stocks, the equal-weight portfolio has produced the lowest return. Among the thirteen sectors studied in this work, the equal-weight portfolio has yielded a higher return for seven sectors. These sectors are banking, consumer durables, financial services, media, pharma, public sector banks, and private banks. The returns yielded by the optimum risk portfolio are higher for the remaining six sectors. The sectors for which the optimum risk portfolio produced a higher return are auto, FMCG, IT, metal, oil & gas, and realty. Another interesting observation is that there both portfolios have yielded negative returns for a substantial number of sectors. While the equal-weight portfolio has produced five sectors, the number of sectors for which the optimum risk portfolio has yielded negative returns is six. For the financial services sector, the equal-weight portfolio produced a low positive return of 5.94%. However, the return yielded by the optimum risk portfolio for the same sector is negative -0.94%. The five sectors for which both portfolios have yielded negative returns are consumer durables, IT, media, pharma, and realty. Hence, from the investors' point of view, these sectors should be avoided as far as possible till the sentiments in the market change. As a final observation, while the performances of the two portfolio design approaches are found to be similar, the equal-weight portfolio outperformed its optimum risk counterpart by a small margin.

**Table 66.** The summary of the results of the portfolio performance

| Sector | EWP Annual Return (%) | MVP/ORP Annual Return (%) |
|---|---|---|
| Auto | 23.52 | 25.78 |
| Banking | 34.43 | 20.25 |
| Consumer Durables | -15.74 | -20.74 |
| Financial Services | 5.94 | -0.94 |
| FMCG | 19.10 | 30.29 |
| IT | -32.09 | -31.16 |
| Media | -6.13 | -17.84 |
| Metal | 14.38 | 41.97 |
| Oil & Gas | 5.78 | 18.46 |
| Pharma | -14.72 | -16.05 |
| Public Sector Banks | 67.70 | 56.34 |
| Private Banks | 22.76 | 14.31 |
| Realty | -13.83 | -11.40 |

## 5. Conclusion

Designing a portfolio for optimizing the return and risk is a very challenging task. In this paper, two approaches to portfolio design are presented. These approaches are the equal-weight portfolio design and the mean-variance portfolio (also known as the optimum risk portfolio) design. Thirteen important sectors listed on the NSE of India are chosen, and the top ten stocks from each sector are identified based on free-float market capitalization. Using the historical stock prices for each stock from Jan 1, 2017, to Dec 31, 2021, two portfolios are designed for each sector. The portfolios are tested over the stock price data from Jan 1, 2022, to Dec 31, 2022. The evaluation of the portfolios is done based on the annual returns yielded by them. It is observed that the returns of the equal-weight portfolio are higher for seven sectors. The sectors for which the equal-weight portfolios have yielded higher returns are banking, consumer durables, financial services, media, pharma, public sector banks, and private banks. However, the optimum risk portfolios produced higher returns than their equal-weight counterparts for six sectors. These sectors are auto, FMCG, IT, metal, oil & gas, and realty. There are five sectors for which both portfolios are found to have produced negative returns over the year 2022. These sectors are consumer durables, IT, media, pharma, and realty. Investors should avoid investing in the stocks of these sectors to avoid loss till the sentiments in the Indian stock market takes a positive turn. Overall, the public sector banks sector yielded the highest return and is found to be the most profitable one. On the other hand, the IT sector has yielded the highest negative return, and hence, this sector looks to be very risky and should be avoided by investors until the Indian stock market sentiments enter into a bull phase. Comparing the performances of the portfolios designed based on other approaches such as hierarchical risk parity (HRP), Eigen portfolios based on principal component analysis, Black-

Litterman portfolio, and hierarchical equal risk contribution (HERC) portfolios on these thirteen sectors and the stocks listed in NIFTY 50 constitutes a future research work.